\documentclass[]{aastex631}

\usepackage{soul}
\usepackage{xcolor}

\shorttitle{X-ray and optical observations of the 2023 outburst in GK Per}
\shortauthors{Kimura et al.}
\graphicspath{{./}{figures/}}

\begin{document}

\title{Evolution of the inner accretion flow and the white-dwarf spin pulse during the 2023 outburst in GK Persei}

\author{Mariko Kimura}

\affiliation{Advanced Research Center for Space Science and Technology, 
College of Science and Engineering, Kanazawa University, \\
Kakuma, Kanazawa, Ishikawa 920-1192, Japan}

\affiliation{Cluster for Pioneering Research, 
Institute of Physical and Chemical Research (RIKEN), \\
2-1 Hirosawa, Wako, Saitama 351-0198, Japan}

\author{Takayuki Hayashi}

\affiliation{Center for Research and Exploration in Space Science and Technology (CRESST II), 
Greenbelt, MD 20771, USA}

\affiliation{Department of Physics, University of Maryland, Baltimore County, 1000 Hilltop Circle, Baltimore, MD 21250, USA}

\affiliation{NASA's Goddard Space Flight Center, X-ray Astrophysics Division, Greenbelt, MD 20771, USA}

\author{Yuuki Wada}

\affiliation{Graduate School of Engineering, Osaka University, 2-1 Yamadaoka, Suita, Osaka, 565-0871, Japan}

\author{Wataru Iwakiri}

\affiliation{International Center for Hadron Astrophysics, 
Chiba University, Chiba 263-8522, Japan}

\author{Shigeyuki Sako}

\affiliation{Institute of Astronomy, Graduate School of Science, The University of Tokyo, 2-21-1 Osawa, Mitaka, Tokyo 181-0015, Japan}

\affiliation{The Collaborative Research Organization for Space Science and Technology, The University of Tokyo, \\
7-3-1Hongo, Bunkyo-ku, Tokyo 113-0033}

\author{Martina Veresvarska}

\author{Simone Scaringi}

\affiliation{Centre for Extragalactic Astronomy, Department of Physics, Durham University, South Road, Durham, DH1 3LE, UK}

\author{Noel Castro-Segura}

\affiliation{Department of Physics, University of Warwick, Gibbet Hill Road, Coventry, CV4 7AL, UK}

\author{Christian Knigge}

\affiliation{Department of Physics and Astronomy. University of Southampton, Southampton SO17 1BJ, UK}

\author{Keith C.~Gendreau}

\author{Zaven Arzoumanian}

\affiliation{Astrophysics Science Division, NASA’s Goddard Space Flight Center, Greenbelt, MD 20771, USA}



\begin{abstract}

We present our X-ray and optical observations performed by NICER, NuSTAR, and Tomo-e Gozen during the 2023 outburst in the intermediate polar GK Persei.
The X-ray spectrum consisted of three components: blackbody emission of several tens of eVs from the irradiated white-dwarf surface, a source possibly including several emission lines around 1 keV, and multi-temperature bremsstrahlung emission from the accretion column.
The 351.3-s white-dwarf spin pulse was detected in X-rays, and the observable X-ray flux from the column drastically decreased at the off-pulse phase, which suggests that the absorption of the column by the accreting gas called the curtain was the major cause of the pulse.
As the system became brighter in optical, the column became fainter, the pulse amplitude became higher, and the energy dependence of pulses became weaker at $<$8~keV.
These phenomena could be explained by the column's more pronounced absorption by the denser curtain as mass accretion rates increased.
The blackbody and line fluxes rapidly decreased at the optical decline, which suggests the expansion of the innermost disk edge with decreasing accretion rates.
The electron scattering or the column geometry may be associated with almost no energy dependence of high-energy pulses.
The irradiated vertically-thick structure at the disk may generate optical QPOs with a period of $\sim$5700~s.

\end{abstract}



\section{Introduction} \label{sec:intro}

Cataclysmic variables (CVs) are close binary systems consisting of a primary white dwarf (WD) and a secondary low-mass star.
The secondary star fills its Roche-lobe, and the gas transferred from the star usually forms an accretion disk around the WD \citep[see][for a comprehensive review]{war95book}.
CVs are classified into two types, magnetic CVs and non-magnetic CVs, by the strength of the WD magnetic field.
Intermediate polars (IPs) are one subclass of magnetic CVs hosting a WD with the magnetic field strength of B $\sim$ 10$^{5-7}$ G.
In these systems, the inner part of the disk is truncated by the WD magnetic field, and the accreting gas channeled to it directly falls onto the WD along field lines at free-fall velocity.
A strong shock is formed, and the gas is heated up to a temperature of the order of $\sim$10~keV.
This high-temperature gas is the source of intense X-ray emission from IPs and is called the accretion column (AC).
Some high-energy X-ray photons from the AC are reflected on the WD surface \citep{muk15reflection,hay18reflection}. 
Some others are absorbed into the WD surface, and a part of it is heated up, which emits soft X-ray photons of about several tens of eVs \citep{eva07IPs,lan09hardXrayCVs}.

The X-ray spin pulse observed in IPs is considered to be produced by the change in the absorption to the observer of the AC by the accreting gas from the disk called the curtain.
Let us consider that we observe an IP system with a magnetic axis inclined to the WD spin axis.
If the AC above the disk equatorial plane points away from the observer, the absorption by the curtain is weak, and the observed X-ray flux is the highest (the light maximum at the on-pulse phase).
If the AC is in front of the WD to the observer, the absorption by the curtain is strong, and the observed X-ray flux is the lowest (the light minimum at the off-pulse phase).
Hereafter, we use the terms ``on-pulse phase'' and ``off-pulse phase'' as the WD spin phase at the light maximum and that at the light minimum to the observer, respectively.
This concept is named the accretion-curtain model \citep{ros88exhyaEXOSAT,hel91aopsc}.

Dwarf novae (DNe), a subgroup of CVs, show sudden brightening of the accretion disk, which is called an outburst.
It is known that several IPs enter dwarf-nova outbursts \citep[see, e.g.,][]{hel97xyariRXTEoutburst,tam22v455and,ish02wzsgeletter}.
DN outbursts are believed to be caused by the thermal-viscous instability in the disk, which is triggered by partial ionization of hydrogen \citep[for reviews, see][]{can93DIreview,osa96review,ham02DImodelproc}.
The thermal equilibrium curve at a given radius of the disk has an unstable branch sandwiched by two stable states: the hot and cool states corresponding to the outburst state with high accretion rates and the quiescent state with low accretion rates, respectively \citep{mey81DNoutburst}.
The disk results in jumping between these two stable states and then, transition waves propagate over the disk \citep{sma84DNoutburst,min85DNDI}.
Although the inner disk is truncated by the WD magnetic field in IPs, the disk instability model is applicable for their outbursts \citep{ham17IPoutburst}.

GK Per is one of IPs with the DN properties \citep{kin79gkperXray}.
This system was discovered by its nova eruption in 1901 and is also known as Nova Persei 1901 \citep{hal1901gkper,pic1901gkper}. 
It is identified as an IP by detecting the 351-s X-ray spin pulse originating from the WD rotation \citep{wat85gkperspin}.
After this eruption, GK Per had kept an almost constant luminosity state of around 13 mag for $\sim$50 yrs and started showing DN outbursts after that.
The outburst in GK Per has a smaller amplitude and longer duration than the typical one, and is characterized by a slow rise.
GK Per is also known as showing quasi-periodic oscillations (QPOs) 
\citep{wat85gkperspin,mor99gkperQPO,nog02gkper,zem17gkper}.
The recent dynamical mass study by \citet{alv21gkper} showed that 
the orbital period ($P_{\rm orb}$), the inclination angle ($i$), the WD mass ($M_1$), and the binary mass ratio ($q \equiv M_2 / M_1$) of GK Per are 1.997~d, 67(5)~deg, 1.03$^{+0.16}_{-0.11}$~M$_{\odot}$, and 0.38(3), respectively.
Here, $M_2$ is the mass of the secondary star.
The distance to the source was measured to be 434$\pm$8~pc by {the \it Gaia} DR3 data \citep{gaia23}.
The orbital period is exceptionally long as a CV because it is mostly less than 12 hrs in the standard evolutionary scenario \citep{RKCat,kni11CVdonor}.

Recently, GK Per entered outbursts every few years and exhibited a new outburst in January 2023.
This outburst was bright at X-ray and optical wavelengths and provided the best opportunity for detailed timing analyses.
We performed observations during this outburst by the X-ray telescope NICER \citep{NICER}, the X-ray satellite NuSTAR \citep{NuSTAR}, and the optical wide-field CMOS camera called Tomo-e Gozen \citep{sak18tomoe}.
We aim to address the following unsolved problems pointed out in previous works: whether the standard disk instability model can explain peculiar outbursts in this system, whether the photoelectric absorption of X-ray photons by the curtain can reproduce all nature of the spin pulse, what complex X-ray emission and QPOs originate from \citep{zem17gkper,pei24gkper,yua16gkper,nog02gkper,mor99gkperQPO,vri05gkper}.
This paper is structured as follows.
Section 2 describes the observation and data reduction. 
Section 3 shows our results of X-ray and optical timing analyses and modelings of X-ray spectra. 
In Section 4, we discuss the outburst mechanism, the X-ray emission source, and the origins of spin pulses and other periodic signals.
We give a summary in Section 5.

\section{Observations and data reduction} \label{sec:obs}

\subsection{NICER}

\begin{table*}[htb]
\caption{Log of observations of GK Per with NICER.  }
\label{nicer-log}
\begin{center}
\begin{tabular}{cccccc}
\hline
NICER~ObsID & ${\rm Start}^{*}$ & ${\rm End}^{*}$ & On-source time$^{\dagger}$ & Average rate$^{\ddagger}$ \\ \hline
  5202530101 & 58.3383 & 58.4635 & 2515 & 11.1 \\ 
  5202530102 & 58.5963 & 58.9138 & 6286 & 23.1 \\ 
  5202530103 & 60.3264 & 60.3333 & 600 & 13.2 \\ 
  5202530104 & 60.6493 & 61.1716 & 995 & 15.5 \\ 
  5202530105 & 62.1349 & 62.4614 & 680 & 13.3 \\ 
  5202530106 & 62.5225 & 63.4288 & 1968 & 16.1 \\ 
  5202530107 & 63.7492 & 64.0740 & 308 & 17.7 \\ 
  5202530108 & 64.9670 & 65.4932 & 1413 & 17.8 \\ 
  5202530109 & 68.3303 & 68.3317 & 121 & 12.3 \\ 
  5202530110 & 68.9740 & 68.9768 & 240 & 19.4 \\ 
  5202530111 & 69.7450 & 70.4604 & 1119 & 13.9 \\ 
  5202530112 & 70.7767 & 71.0409 & 811 & 15.8 \\ 
  5202530113 & 71.7446 & 72.3288 & 1304 & 18.5 \\ 
  5202530114 & 72.5831 & 73.4919 & 2056 & 16.2 \\ 
  5202530115 & 73.6161 & 74.2637 & 868 & 16.0 \\ 
  5202530116 & 74.5196 & 74.9732 & 639 & 13.5 \\ 
  5202530117 & 75.6795 & 76.1342 & 860 & 15.4 \\ 
  5202530118 & 77.8059 & 77.8780 & 1329 & 15.5 \\ 
  5202530119 & 80.3191 & 80.3339 & 1177 & 16.1 \\ 
  5202530120 & 80.9633 & 80.9711 & 675 & 12.9 \\ 
  5202530121 & 81.9367 & 82.1347 & 2002 & 16.8 \\ 
  5202530122 & 83.2265 & 83.2322 & 500 & 9.2 \\ 
  5202530123 & 83.5495 & 84.2037 & 4435 & 14.0 \\ 
  5202530124 & 84.6403 & 84.7231 & 4014 & 11.9 \\ 
  5202530125 & 85.6725 & 85.6991 & 2305 & 17.4 \\ 
  5202530126 & 86.6447 & 87.1731 & 2631 & 15.4 \\ 
  5202530127 & 87.6719 & 88.2050 & 2951 & 19.0 \\ 
  5202530128 & 88.9092 & 89.4339 & 1853 & 17.3 \\ 
  5202530129 & 89.6920 & 89.8111 & 1336 & 13.8 \\ 
  5202530130 & 91.5416 & 91.6942 & 3776 & 17.1 \\ 
  5202530131 & 93.7371 & 94.0717 & 3878 & 11.8 \\ 
  5202530132 & 94.9593 & 95.0356 & 2456 & 12.7 \\ 
  5202530133 & 96.9594 & 97.0412 & 3007 & 13.1 \\ 
  5202530134 & 97.9999 & 98.0728 & 1402 & 12.2 \\ 
  5202530135 & 98.9675 & 99.0329 & 281 & 10.2 \\ 
\hline
\multicolumn{5}{l}{$^{*}$BJD$-$2459900.0.}\\
\multicolumn{5}{l}{$^{\dagger}$Units of seconds.}\\
\multicolumn{5}{l}{$^{\ddagger}$NICER count rate in 0.3--7~keV in units of counts/sec.}\\
\end{tabular}
\end{center}
\end{table*}

Hereafter, all observation times are converted to barycentric Julian date (BJD).
The X-ray telescope NICER onboard the International Space Station (ISS) monitored GK Per during its 2023 outburst, starting 
from BJD 2459958.
The NICER effective area around 1~keV is larger than the Swift one by $\sim$10 times.
The observation IDs (ObsIDs) of the monitoring are given in Table \ref{nicer-log}. 
This work used HEAsoft version 6.31.1 including FTOOLS\footnote{$<$https://heasarc.gsfc.nasa.gov/ftools/$>$} \citep{FTOOLS} for data reduction and analyses.
The data were reprocessed with the pipeline tool \texttt{nicerl2}, which used the NICER Calibration Database (CALDB) version later than 2022 October 31, for producing light curves and time-averaged spectra. 
NICER is composed of 56 modules of silicon drift detectors (SDDs), 52 of which are operating in orbit, including two noisy modules, IDs 14 and 34. 
In our data reduction process with \texttt{nicerl2}, we filtered out the data of these two modules. 
The light curves were generated by \texttt{lcurve}.
The source and background spectra were extracted with \texttt{nibackgen3C50} version 7. 
For spectral analyses, we obtained the response matrix file and the ancillary response file for a specific set of 50 detectors to match the default settings of the background model.\footnote{The method is described at $<$https://heasarc.gsfc.nasa.gov/docs/nicer/analysis\_threads/arf-rmf/$>$, and we use the additional data version xti20200722.}

\subsection{NuSTAR}

\begin{table*}[htb]
\caption{Log of observations of GK Per with NuSTAR.  }
\label{nustar-log}
\begin{center}
\begin{tabular}{cccccc}
\hline
NICER~ObsID & ${\rm Start}^{*}$ & ${\rm End}^{*}$ & On-source time$^{\dagger}$ & Average rate$^{\ddagger}$ \\ \hline
  90901305002 & 97.6959 & 98.6047 & 43191 & 1.18 \\ 
\hline
\multicolumn{5}{l}{$^{*}$BJD$-$2459900.0.}\\
\multicolumn{5}{l}{$^{\dagger}$Units of seconds.}\\
\multicolumn{5}{l}{$^{\ddagger}$NuSTAR count rate in 3--79~keV in units of counts/sec.}\\
\end{tabular}
\end{center}
\end{table*}

The NuSTAR Target of Opportunity (ToO) observations were carried out on 2023 February 22 (ObsID: 90901305002). 
The observation log is given in Table \ref{nustar-log}. 
The observation time was partially overlapped with that of the NICER observation of ObsID 5202530134 on the same day. 
The data were reprocessed with \texttt{nupipeline} and the NuSTAR CALDB as of 2022 May 10.  
The light curves, time-averaged spectra, and response and ancillary response files were obtained with \texttt{nuproducts}.  
The background region was circular with a radius of 100'' at a blank sky area.  
We determined the circular source region centered at the target position with a 50'' radius.

\subsection{Tomo-e Gozen}

\begin{table*}[htb]
\caption{Log of observations of GK Per with Tomo-e Gozen.  }
\label{tomoe-log}
\begin{center}
\begin{tabular}{cccccc}
\hline
Date & ${\rm Start}^{*}$ & ${\rm End}^{*}$ & On-source time$^{\dagger}$ & Average magnitude$^{\ddagger}$ \\ \hline
2023-02-17 & 92.9807 & 93.0407 & 2854 & 12.0 \\ 
2023-02-21 & 96.9583 & 97.0439 & 3821 & 12.1 \\ 
2023-02-22 & 98.0550 & 98.0676 & 1090 & 12.3 \\ 
2023-02-23 & 98.9521 & 99.0357 & 3821 & 12.3 \\ 
\hline
\multicolumn{5}{l}{$^{*}$BJD$-$2459900.0.}\\
\multicolumn{5}{l}{$^{\dagger}$Units of seconds.}\\
\multicolumn{5}{l}{$^{\ddagger}$Magnitude without a filter. The sensitive wavelengths are 380--710 nm.}\\
\end{tabular}
\end{center}
\end{table*}

Tomo-e Gozen is an optical wide-field video observation system composed of 84 chips of CMOS image sensors on the 1.05 m Kiso Schmidt telescope \citep{sak18tomoe}.
This system is capable of obtaining consecutive frames with timestamps of 0.2-ms absolute accuracy. 
The frame rate is increased by reading a part of the pixels of the image sensor.
We observed GK Per with 19 frames per second (fps).
The observation log is given in Table \ref{tomoe-log}.

After the dark subtraction and flat fielding for the raw data, 
we extracted the light curve by using the `Photutils' python package \citep{photutils}. 
We stacked every ten frames to 0.5-s exposure, searched bright stars in each stacked frame, and inquired their $G$-band magnitudes from Gaia EDR3 catalog \citep{gaia21}.
The background was subtracted by `SExtractor'.\footnote{$<$https://www.astromatic.net/software/sextractor/$>$}
We made the diagram of the $G$ magnitude vs.~the measured count for several bright stars and estimated the zero count $C_{\rm zero}$ corresponding to 0 mag.
We performed the relative photometry for the target by using the $C_{\rm zero}$ curve.
Since Tomo-e Gozen is sensitive to rapid time variations of the sky condition, we need to perform relative photometry for the target with several reference stars in the same image to check whether the sky condition affects the target light curve or not.
If the sky condition is bad, $C_{\rm zero}$ fluctuates.
We removed these outlier points.

\subsection{AAVSO data}

We also used optical data obtained by the American Association of Variable Star Observers (AAVSO).
The AAVSO archive data are available at this URL\footnote{$<$http://www.aavso.org/data/download/$>$.}.
Each observer performed the relative photometry by using nearby comparison stars.
The magnitude of each comparison star was measured by the AAVSO Photometric All-Sky Survey \citep[APASS: ][]{APASS} from the AAVSO Variable Star Database\footnote{$<$http://www.aavso.org/vsp/$>$.}.

\section{Results}

\subsection{Overall outburst behavior} \label{sec:overall}

\begin{figure*}[htb]
\epsscale{0.8}
\plotone{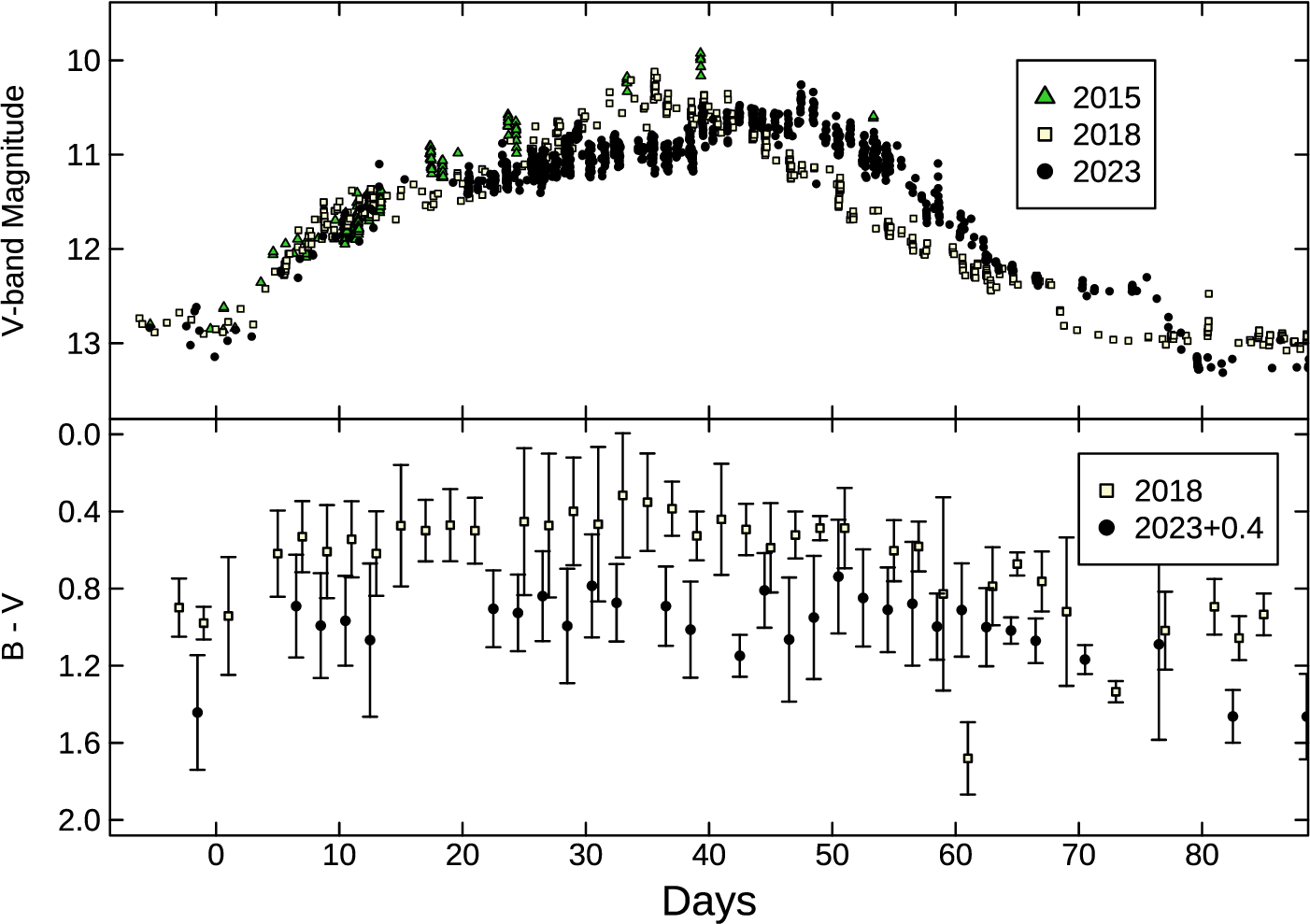}
\caption{
Overall optical $V$-band light curves and $B-V$ color evolution during outbursts in GK Per.  
The data were provided by the AAVSO.
The upper panel shows light curves during the 2015, 2018, and 2023 outbursts. 
The lower panel shows $B-V$ color evolution during the 2018 and 2023 outbursts.
The triangle, rectangle, and circle represent the data of the 2015, 2018, 2023 outbursts, respectively.
}
\label{overall-optical}
\end{figure*}

The optical light curve and the $B-V$ color evolution during the 2023 outburst are exhibited in Figure \ref{overall-optical} in comparison with those in previous outbursts.
The outburst light curve seems to be composed of three parts: the early phase of the rising part (a steeper rise) during Day 0--15, the later phase of the rising part (a slower rise) during Day 15--48, and a rapid decline during Day 48--66.
The rising or fading rates of these three parts are $-$9.6, $-$26.1, and 9.0~d~mag$^{-1}$, respectively.
The $B-V$ color became bluest around the outburst maximum.

\begin{figure*}[htb]
\epsscale{0.8}
\plotone{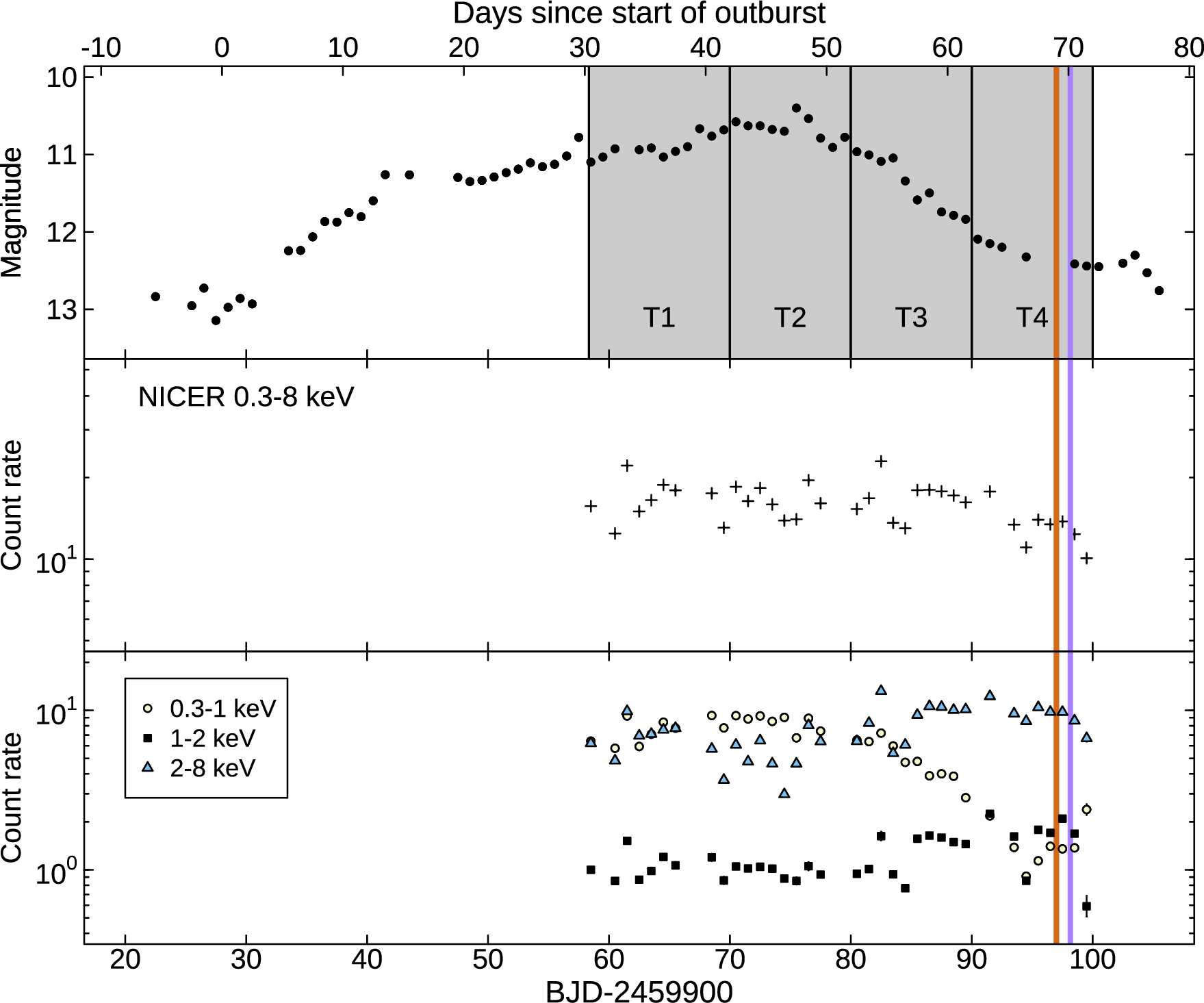}
\caption{
Overall optical $V$-band light curves (top panel), NICER X-ray light curves in the 0.3--8~keV (middle panel), 0.3--1~keV, 1--2~keV, and 2--8~keV bands (bottom panel) during the 2023 outburst in GK Per.
In the bottom panel, the white circle, black rectangle, and blue triangle represent 0.3--1~keV, 1--2~keV, and 2--8~keV light curves.
The data are averaged per day.
The orange and purple lines indicate the time of simultaneous observations by NICER and Tomo-e-Gozen, and by NICER and NuSTAR, respectively.
Here, T1, T2, T3, and T4 represent the time zones before BJD 2459970, BJD 2459970--2459980, BJD 2459980–-2459990, and BJD 2459900–2460000, respectively.
We divide the data into these time zones in section \ref{sec:res-spin-pulse} in order to explore the time evolution of WD spin pulses.
}
\label{overall}
\end{figure*}

We compared the optical and X-ray light curves during the 2023 outburst in Figure \ref{overall}.
The X-ray flux in the 0.3--8 keV seemed constant during this outburst (see the middle panel of Figure \ref{overall}), which may be similar to that in the 2018 outburst \citep{pei24gkper}.
The 0.3--1~keV flux basically traced the optical light curve (see white circles in the bottom panel).
On the other hand, the 2--8~keV light curve was anti-correlated with the optical light curve (see blue triangles in the same panel).
The 1--2~keV flux increased since the middle of the fading stage (see black rectangles in the same panel).

\subsection{Broadband X-ray spectral analyses} \label{sec:sed}

\begin{figure*}[htb]
\epsscale{0.7}
\plotone{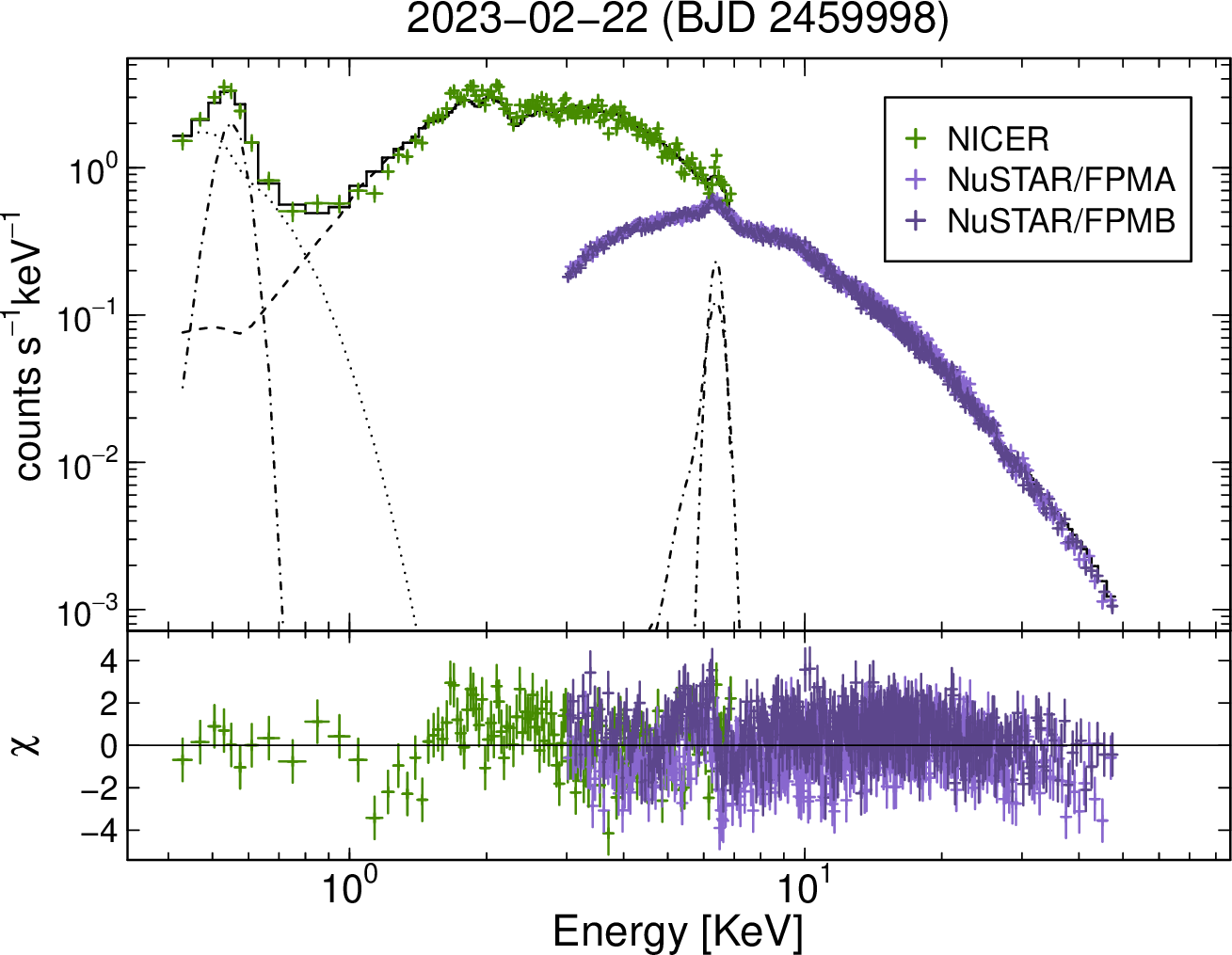}
\caption{
Broad-band X-ray spectrum of GK Per on BJD 2459998 (2023 Feburuary 22), overlaid with the best-fit spectral model of \texttt{Tbabs*(bbody + gaussian + pwab*(reflect*vmcflow + gaussian))}. 
The green crosses represent the NICER data. 
The purple and dark purple crosses represent the NuSTAR FPMA and FPMB data, respectively. 
The dot, dashed, and dot-dashed lines represent the best-fit model components of blackbody, cooling-flow, and emission lines, respectively.
The solid line shows the total best-fit model emission. 
The normalization of the NuSTAR SED was 0.98 of the NICER SED.
}
\label{spec230222}
\end{figure*}

\begin{table*}[htb]
    \caption{
    Best-fit parameters for models of \texttt{Tbabs*(bbody + gaussian + pwab*(reflect*vmcflow + gaussian))} in the simultaneous spectral model-fitting of the simultaneous NICER and NuSTAR observation data of GK Per on BJD 2459998 (2023\ February 22).  
    The errors represent 90\% confidence ranges. 
    Gaussians 1 and 2 denote oxygen and iron fluorescence lines, respectively. 
    }
    \vspace{2mm}
    \label{parameter-230222}
    \centering
\begin{tabular}{ccc}
\hline
Model & Parameter & Best-fit value \\
\hline
Tbabs & $N_{\rm H}$$^{*}$ 
      & 0.16 (fixed) \\
\hline
bbody & $T_{\rm BB}$$^{\dagger}$ 
      & 83$^{+5.7}_{-5.3}$ \\
      & $L_{\rm BB}$$^{\ddagger}$ 
      & 0.56$^{+0.15}_{-0.16}$ \\
\hline
pwab & $N_{\rm H, min}$$^{*}$
      & 0.34$\pm$0.01 \\
      & $N_{\rm H, max}$$^{*}$
      & 10.7$^{+0.46}_{-0.75}$ \\
\hline
reflect & $\Omega / 2\pi$$^{\P}$
		& 1.0 (fixed) \\
		& $\cos i$$^{\S}$ 
		& 0.39 (fixed) \\
\hline
vmcflow & $kT_{1}$$^{\|}$
		& 54.1$^{+2.3}_{-0.6}$ \\
		& $Z_{\rm O}$$^{**}$ 
        & 0.48$^{+0.06}_{-0.07}$ \\
        & $\dot{M}_{\rm acc}$$^{\dagger\dagger}$
        & 6.4$^{+0.2}_{-0.1}$$\times$10$^{-10}$ \\
\hline
gaussian 1 
& $E_1$$^{\ddagger\ddagger}$ 
		& 0.53 (fixed) \\
		& $\sigma_1$$^{\P\P}$
        & 0.001 (fixed) \\
    	& Norm$_1^{\S\S}$ 
        & (6.1$\pm$1.1)$\times$10$^{-4}$ \\
\hline
gaussian 2 & $E_2$$^{\ddagger\ddagger}$ 
		& 6.4 (fixed) \\
		& $\sigma_2$$^{\P\P}$
        & 0.18$\pm$0.02 \\
        & Norm$_2^{\S\S}$ 
        & (3.3$^{+0.2}_{-0.1}$)$\times$10$^{-4}$ \\
\hline
$\chi^2$/dof & & 1.54 \\
\hline
\multicolumn{3}{l}{\parbox{160pt}{$^{*}$Equivalent hydrogen column in 10$^{22}$ atoms cm$^{-2}$.}}\\
\multicolumn{3}{l}{$^{\dagger}$Blackbody temperature in eV.}\\
\multicolumn{3}{l}{\parbox{160pt}{$^{\ddagger}$Blackbody luminosity, $L_{36}/D_{10}^2$, where $L_{36}$ is the source luminosity in units of $10^{36}$ erg/s and $D_{10}$ is the distance to the source in units of 10 kpc.}}\\
\multicolumn{3}{l}{$^{\P}$Reflection scaling factor.}\\
\multicolumn{3}{l}{\parbox{160pt}{$^{\S}$Here, $i$ denotes the inclination angle of the system, and we applied $i$ = 67 deg.}}\\
\multicolumn{3}{l}{$^{\|}$Plasma temperature in keV.}\\
\multicolumn{3}{l}{\parbox{160pt}{$^{**}$Oxygen abundance with respect to the solar one.}}\\
\multicolumn{3}{l}{\parbox{160pt}{$^{\dagger\dagger}$Mass accretion rate in units of $\dot{M}_{\odot}$~yr$^{-1}$.}}\\
\multicolumn{3}{l}{$^{\ddagger\ddagger}$Line energy in keV.}\\
\multicolumn{3}{l}{$^{\P\P}$Line width in keV.}\\
\multicolumn{3}{l}{$^{\S\S}$Total photons/cm$^2$/s in the line.}\\
\end{tabular}
\end{table*}

We obtained partially overlapped NICER and NuSTAR observations on BJD 2459998 (see also the purple line in Figure \ref{overall}).
We cut the NICER spectrum including the data taken during ISS daytime below 0.4~keV because of high background rates.
Empirically, the X-ray spectrum from IPs has been modeled by a cooling-flow model, which is a kind of multi-temperature thin thermal plasma model \citep[e.g.,][]{hay11v1223sgr,dob17mvlyr,tsu18gammaCas}.
We fitted the spectra with the model \texttt{Tbabs*(bbody + gaussian + pwab*(reflect*vmcflow + gaussian))} in the \texttt{XSPEC} software \citep{XSPEC}, where \texttt{Tbabs}, \texttt{pwab}, \texttt{bbody}, \texttt{reflect}, and \texttt{vmcflow} denote X-ray absorption by the interstellar medium, partial X-ray absorption by intrinsic medium to the object, blackbody radiation, a convolution model for reflection from neutral material, and a cooling flow model, respectively.
The first and second \texttt{gaussian} models represent oxygen fluorescence K$\alpha$ line at 0.53~keV and iron fluorescence K$\alpha$ line at 6.4~keV, respectively.
The line center energies of these models are fixed.
As for the former emission line, the line width parameter $\sigma$ is fixed at 0.001 keV because of the limitation of NICER energy resolution.

We fit eight NICER spectra averaged per five days by tying the column-density parameter $N_{\rm H}$ in the \texttt{Tbabs} model, and estimated it as 1.6$\times$10$^{21}$~cm$^{-2}$ (see also section \ref{sec:evolution-xray-spec}), which is consistent with the value reported by past works \citep{zem17gkper,pei24gkper}.
Hereafter, we fixed $N_{\rm H}$ in that model at this value.
Also, we fixed the solid angle of the reflector subtending the hot plasma at 1, which corresponds to the parameter $\Omega / 2\pi$ in the \texttt{reflect} model, by assuming the reflector is the WD.
Since the accretion rate onto the WD in GK Per is high during outburst, the AC height is low as $<$10\% of the WD radius \citep{sul16gkper,wad18gkper}.
We tied the abundance $Z$ in the \texttt{reflect} model and the oxygen abundance $Z_{\rm O}$ in the \texttt{vmcflow} model.
The iron and nickel abundances were fixed at 0.105 and 0.1, respectively, which were estimated by \citet{zem17gkper}.
The best-fit model parameters are summarized in Table \ref{parameter-230222}.

We found that the blackbody (BB) emission was dominant at less than 1 keV.
The best-fit temperature and radius were 78~eV and 5.3$\times$10$^{5}$~cm, respectively.
The irradiated region was only 0.0001\% of the WD surface, and the temperature of the irradiated WD surface was much higher than the typical temperature of non-irradiated WDs in CVs during outbursts, which is $\sim$40,000~K (see table 2.8 in \citet{war95book} and \citet{god17ugem}).
Here, we assume that the radius of $\sim$1-$M_{\odot}$ WD in GK Per is 5$\times$10$^{8}$~cm \citep{nau72WDmassradius,alv21gkper}.
The optically-thin and multi-temperature bremsstrahlung emission was dominant above 1 keV.
The maximum temperature was 54~keV and the mass accretion rate was 6.4$\times$10$^{-10}$~$M_{\odot}$~yr$^{-1}$.
These values are close to previously reported parameter values in the quiescent state \citep{zem17gkper,wad18gkper,pei24gkper}.
This would be because this spectrum was taken at the end of the outburst.
The unabsorbed flux in the 0.3--50 keV was 5.0$\times$10$^{-10}$~ergs~s$^{-1}$cm$^{-2}$, which corresponds to the X-ray luminosity of 1.1$\times$10$^{34}$~ergs~s$^{-1}$.

\subsection{Evolution of X-ray spectra} \label{sec:evolution-xray-spec}

\begin{figure*}[htb]
\epsscale{1.0}
\begin{minipage}{0.49\hsize}
\plotone{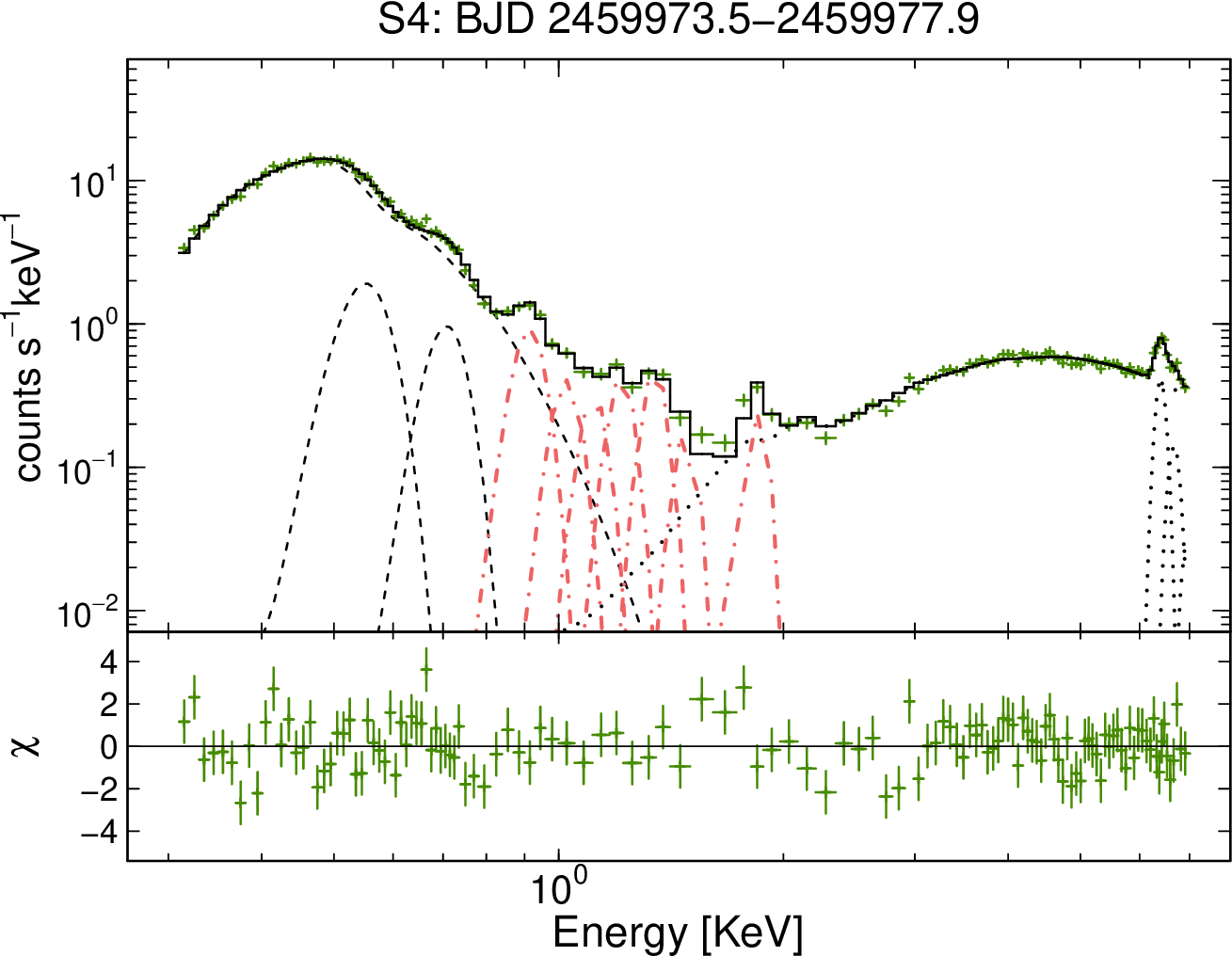}
\end{minipage}
\begin{minipage}{0.49\hsize}
\plotone{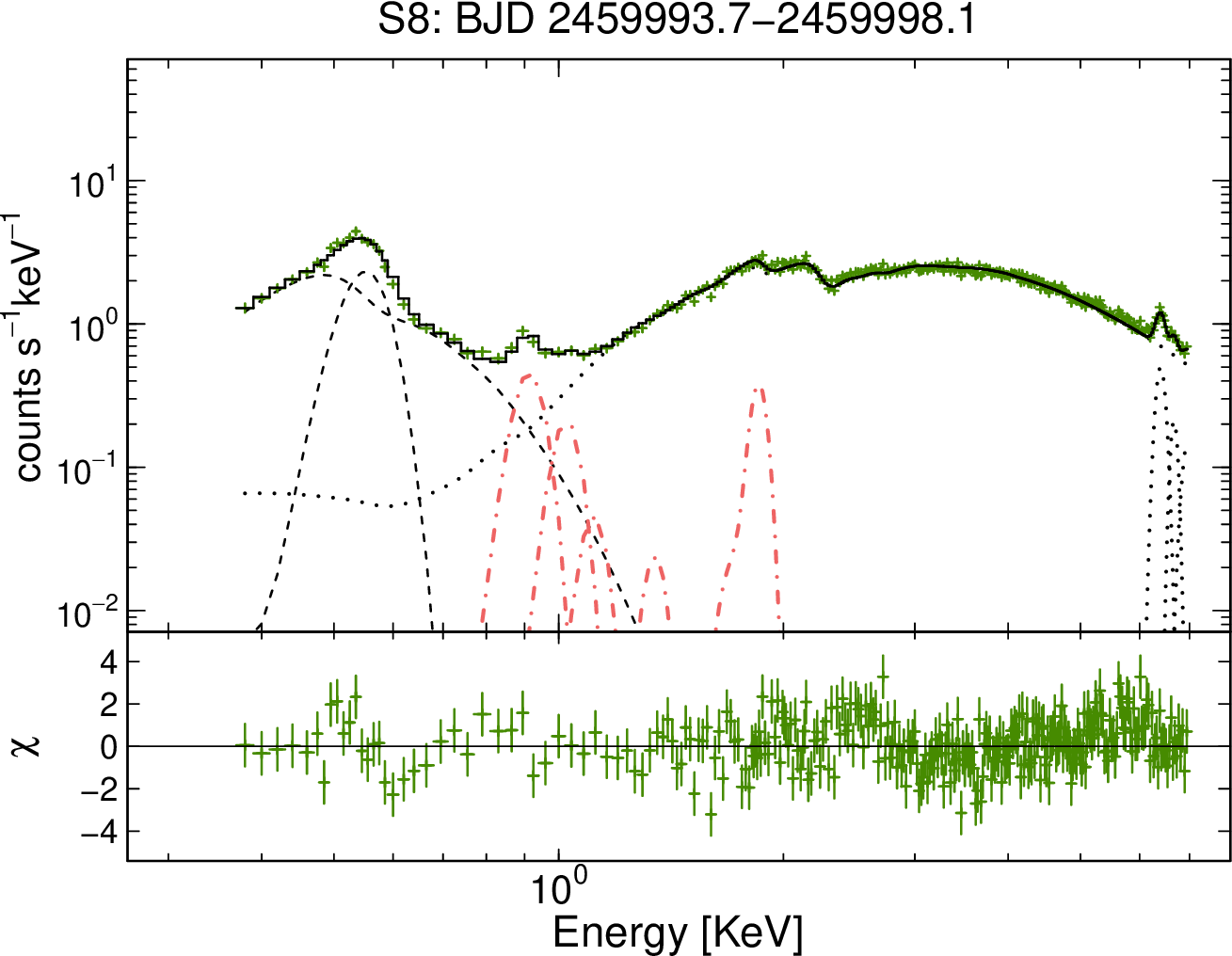}
\end{minipage}
\caption{
X-ray spectra of GK Per during the 2023 outburst, which are overlaid with the best-fit spectral model of \texttt{Tbabs*(gaussian + gaussian + bbody + gaussian + gaussian + gaussian + gaussian + gaussian + gaussian + gaussian + pwab*(reflect*powerlaw + gaussian + gaussian + gaussian))}. 
The left and right panels show the averaged NICER spectra with their best-fit models in S4 and S8, respectively.
The green cross represents the averaged NICER spectrum. 
The red dot-dashed line denotes the best-fit models of neon, iron, magnesium, and silicon emission lines.
The dashed line represents the best-fit model of blackbody emission with oxygen and iron lines.
The dot line represents the best-fit model components of hot plasma plus three iron lines.
The solid line shows the total best-fit model emission. 
}
\label{spec-5d-04-08}
\end{figure*}

\begin{figure*}[htb]
\epsscale{0.5}
\plotone{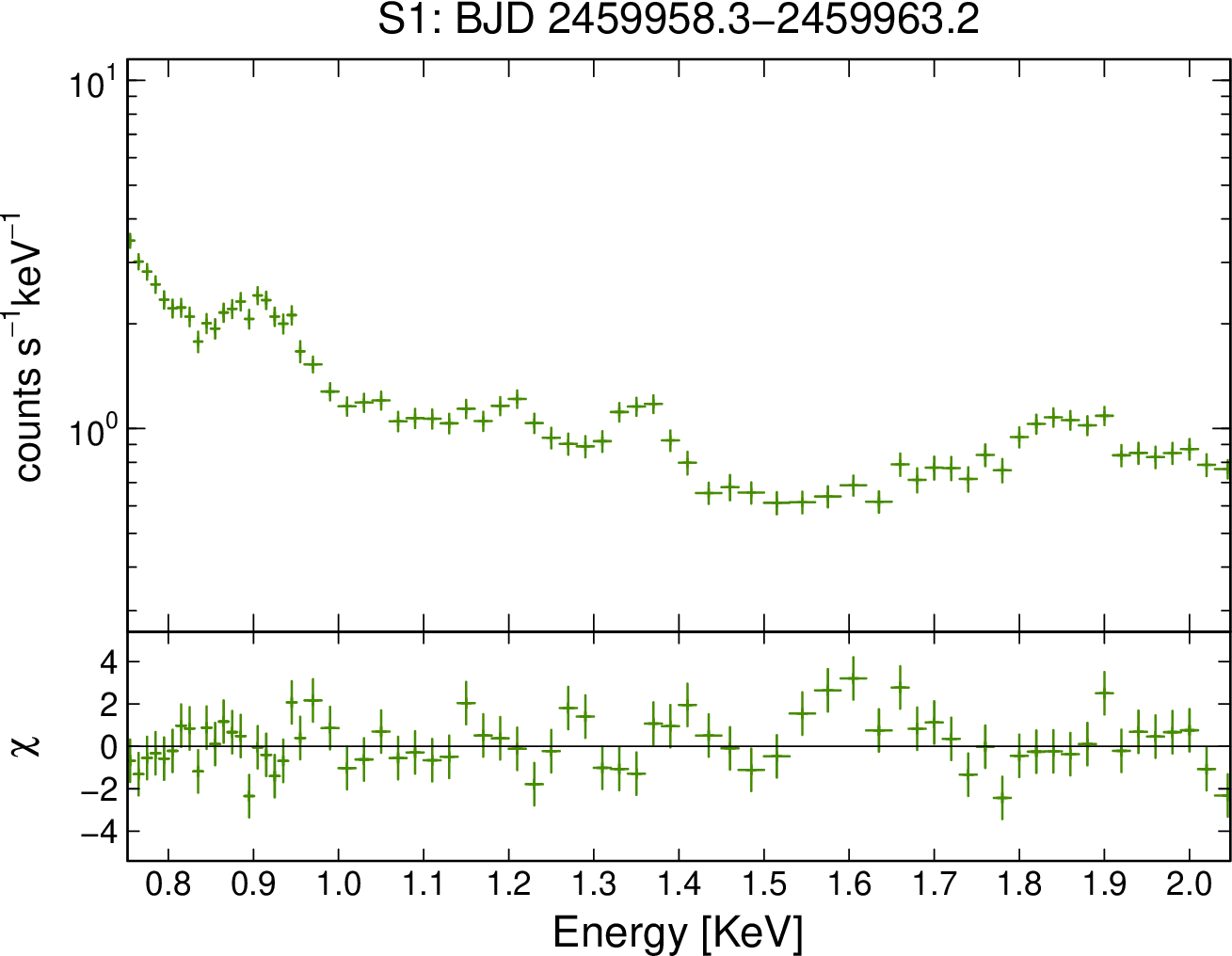}
\caption{
An enlarged figure around 1 keV in a linear energy axis for the spectrum in S1.
}
\label{linear-spec}
\end{figure*}

\begin{figure*}[htb]
\epsscale{0.8}
\plotone{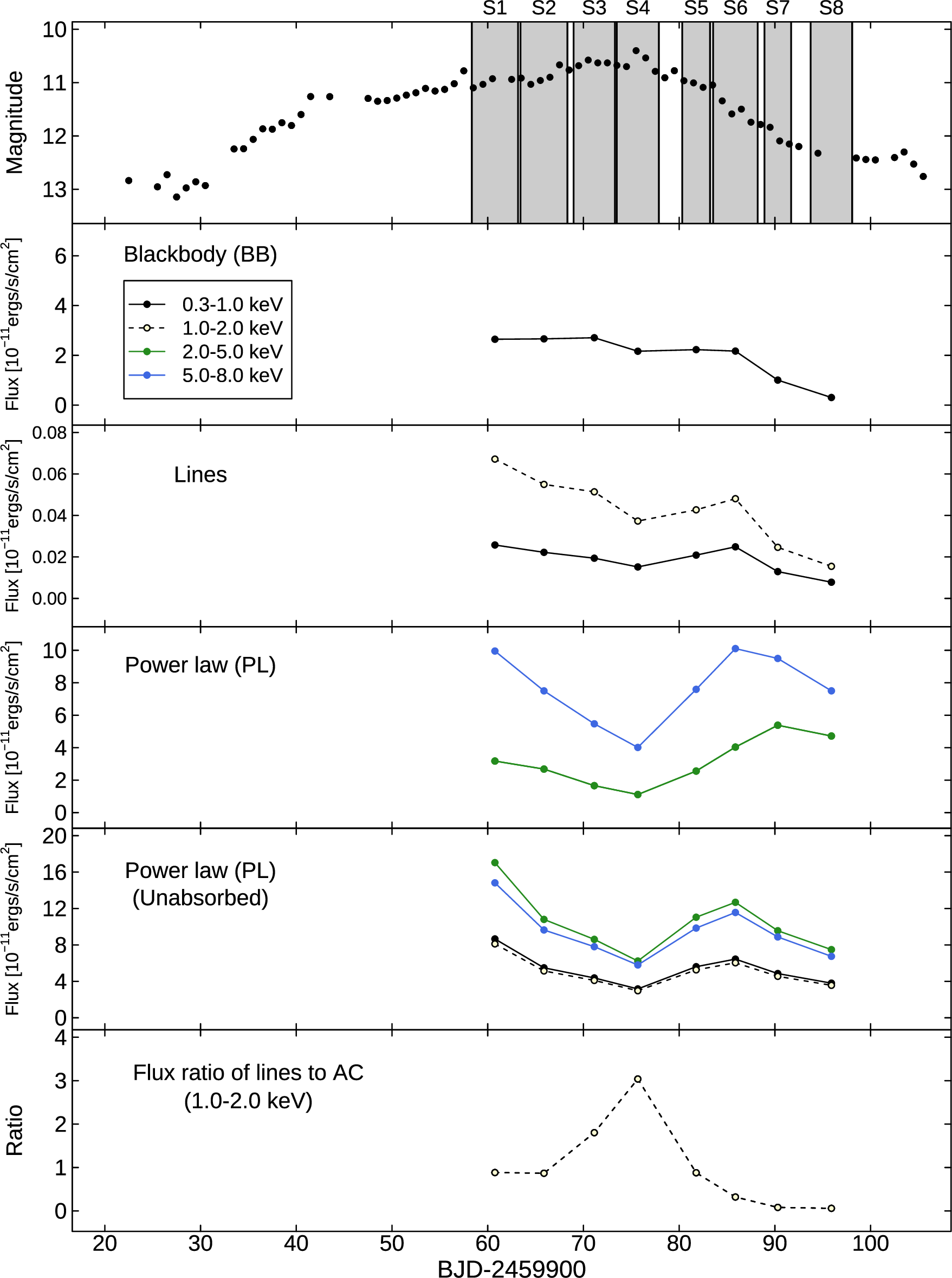}
\caption{
Optical light curves (top panel), X-ray fluxes of the best-fit model components in the 0.3--1 keV, 1--2 keV, 2--5 keV, and 5--8 keV bands (second, third, fourth, and fifth panels), and the flux ratio of neon, magnesium, iron, and silicon emission lines to the sum of the power law and three Gaussian components in the 1--2~keV band (the bottom panel). Here we omit some model flux curves if they have a little contribution.
S1, S2, S3, S4, S5, S6, S7, and S8 represent the time zones BJD 2459958.3--2459963.2, BJD 2459963.4--2459968.3, BJD 2459969.0--2459973.3, BJD 2459973.5--2459977.9, BJD 2459980.3--2459983.2, BJD 2459983.5--2459988.2, BJD 2459988.9--2459991.7, BJD 2459993.7--2459998.1, respectively.
We divide the data into these time zones in order to investigate the time evolution of X-ray spectra.
}
\label{modelflux}
\end{figure*}

We also tried to model the NICER spectra averaged per five days to investigate the time evolution of each spectral component.
Eight spectra were obtained and each spectrum is numbered as S1, S2, S3, S4, S5, S6, S7, and S8, respectively, from the earliest to the latest.
These time intervals are denoted in the top panel of Figure \ref{modelflux}.
We needed to add neon, iron, magnesium, and silicon emission lines to the model, which were prominent in the bright state during previous outbursts \citep{vri05gkper,zem17gkper}.
The line width was comparable with the energy resolution ($\sim$0.1~keV around 1~keV), and the detailed structure, e.g., the He-like triplet, was not resolved (see Figure \ref{linear-spec}).

We modeled the averaged NICER spectra by the model \texttt{Tbabs*(gaussian + gaussian + bbody + gaussian + gaussian + gaussian + gaussian + gaussian + gaussian + gaussian + pwab*(reflect*powerlaw + gaussian + gaussian + gaussian))}.
The \texttt{powerlaw} model is a simple photon power-law model.
We added this model instead of the \texttt{vmcflow} model since it is hard to determine $T_{\rm max}$ only with the NICER spectrum below 7~keV.
Since we aimed to investigate the change in the absorption effect, we fixed the photon index at 1.48 for all spectra.
This value was derived from the modeling with the power-law model of the best-fit \texttt{vmcflow} model for the simultaneously observed NICER and NuSTAR spectrum presented in Figure \ref{spec230222} and Table \ref{parameter-230222}.
We confirmed that this slope does not change if $T_{\rm max}$ is higher than 10 keV.
This condition was satisfied in previous outbursts of GK Per \citep{yua16gkper,wad18gkper,pei24gkper}.
The abundance $Z$ in the \texttt{reflect} model was fixed at the best-fit $Z_{\rm O}$ value obtained in section \ref{sec:sed}.

The line center energies of the first two Gaussian models are 0.53~keV (fixed) and 0.705~keV (fixed), respectively.
They represent the oxygen fluorescence K$\alpha$ line and the iron fluorescence L$\alpha$ line.
The next seven Gaussian models represent the emission lines of Ne IX (0.91 keV), Ne X (1.02 keV), Fe XXIV (1.11 keV), Ne X (1.21 keV), Mg XII (1.34 keV), Mg XI (1.47 keV), and Si XIII (1.85 keV).
The line width parameter $\sigma$ is fixed at 0.001~keV.
The last three Gaussian models represent the three iron lines (Fe fluorescence line at 6.4 keV, Fe He-like complex at 6.67 keV, and Fe H-like line at 6.97 keV), and the line center energies are fixed.

Figure \ref{spec-5d-04-08} shows examples of the averaged spectra and their models.
We can see the emission lines, in particular, highly-ionized ones denoted by red color, were dominant around 1~keV at the bright state, and were weaker and weaker during the fading stage of this outburst.
Figure \ref{modelflux} displays the unabsorbed flux of the BB component, that of the sum of seven emission lines of neon, iron, magnesium, and silicon, and that of the power-law (PL) component plus three iron lines in four energy bands.
Here, the ``unabsorbed'' flux means the flux without the interstellar absorption, i.e., $N_{\rm H}$ is zero in the model \texttt{Tbabs}.
The intrinsic absorption represented by the model \texttt{pwab} is still multiplied by the PL and three Gaussian components for the multi-temperature plasma and iron emission lines in the fourth panel of Figure \ref{modelflux}. 
We also extracted the unabsorbed flux of these components by removing the \texttt{pwab} model (see the fifth panel of the same figure).
The blackbody component was dominant at $<$1 keV.
The total flux of seven Gaussian models for neon, magnesium, and silicon lines was dominant at 1--2 keV, though its 0.3--1~keV flux became comparable at the end of this outburst.
The total flux of the PL plus three Gaussian models was dominated above 2 keV.
Although the line flux overwhelmed the flux of the PL component in the 1--2 keV band till the middle of the fading stage, this relation reversed after S6.
The PL flux significantly decreased around the outburst maximum.
The BB and line fluxes rapidly faded after S6.
The PL component was the softest near the outburst maximum.

\subsection{Period analyses} \label{sec:period}

\begin{figure*}[htb]
\begin{minipage}{0.33\hsize}
\plotone{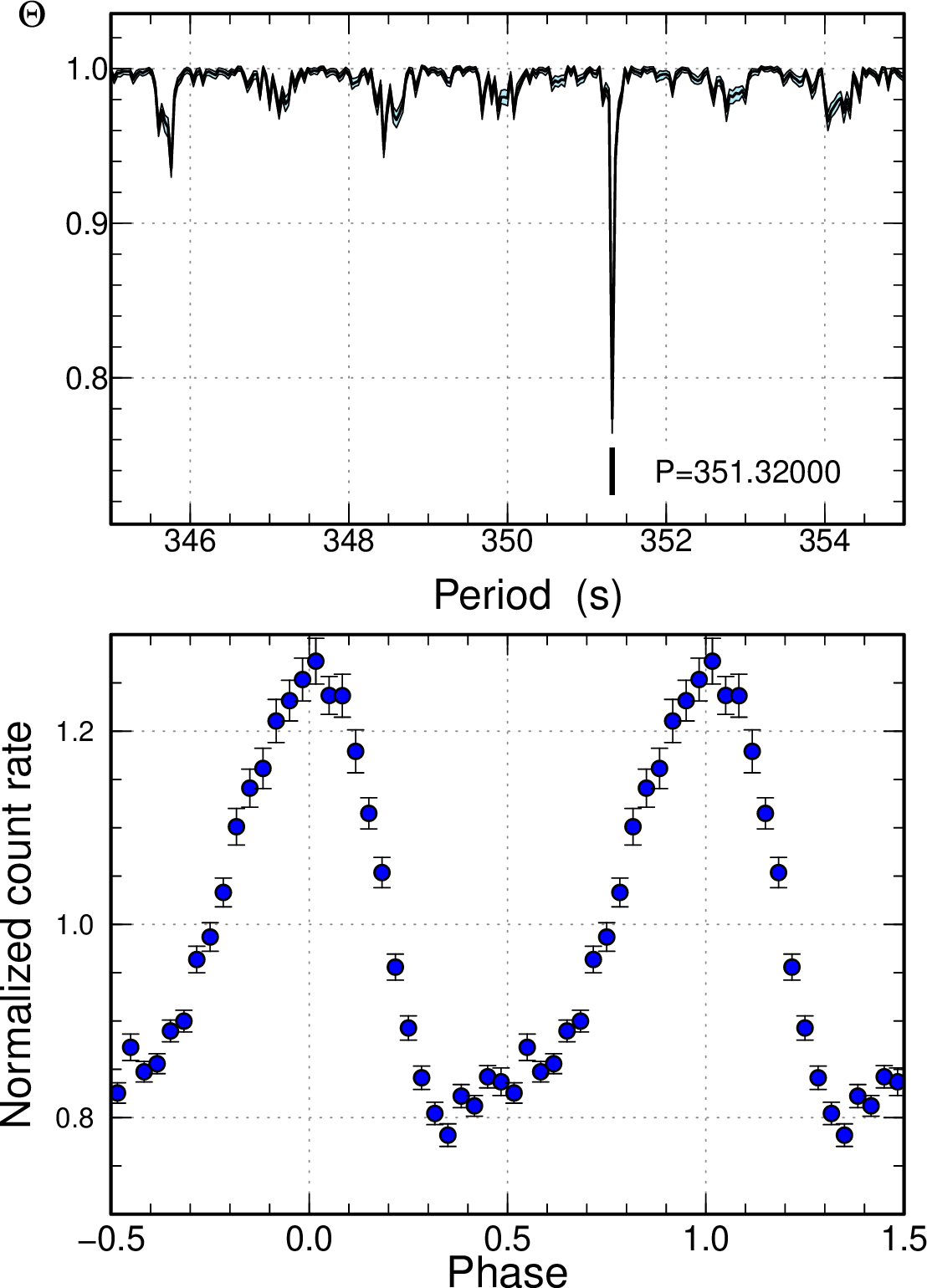}
\end{minipage}
\begin{minipage}{0.33\hsize}
\plotone{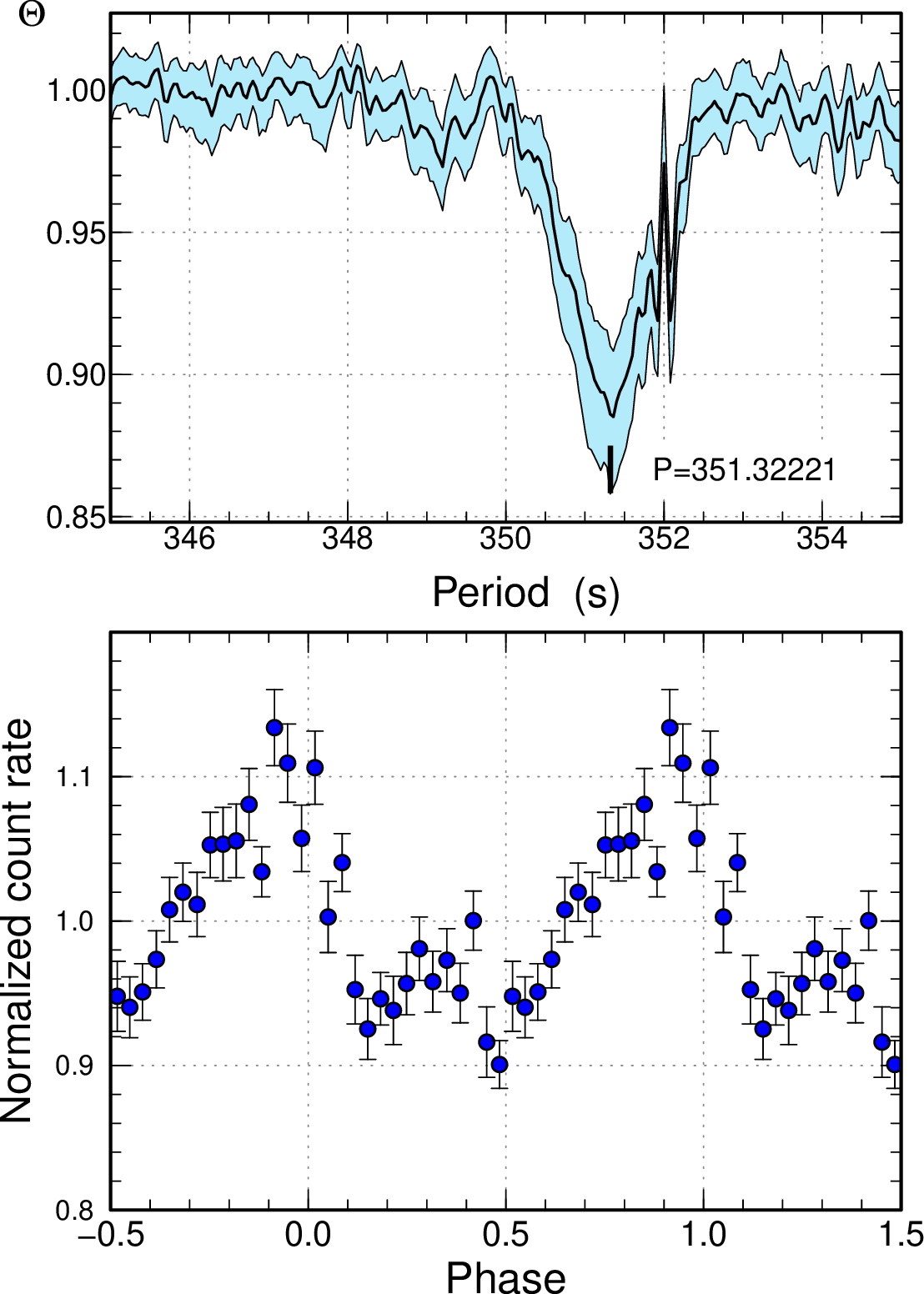}
\end{minipage}
\begin{minipage}{0.33\hsize}
\plotone{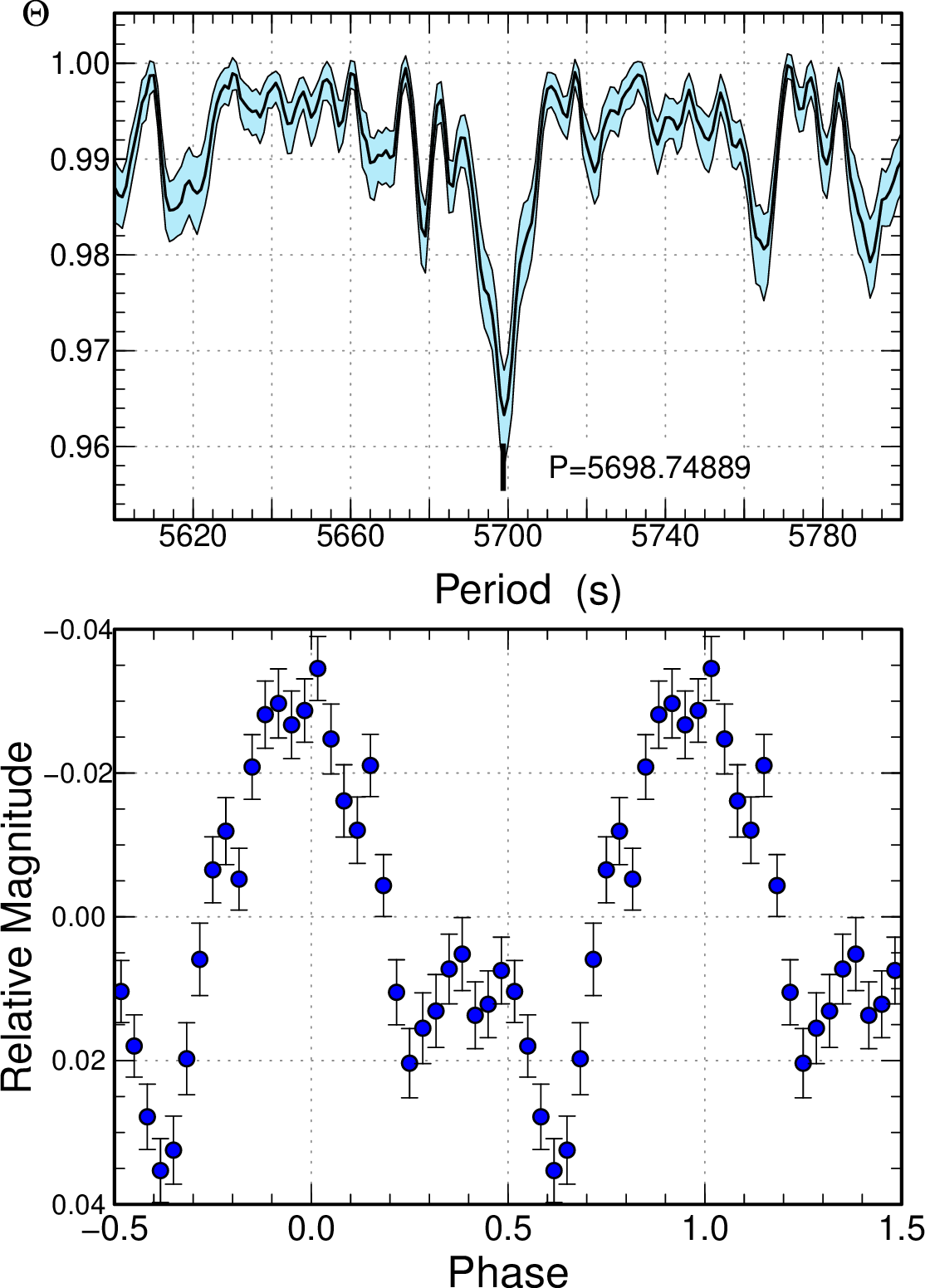}
\end{minipage}
\caption{
PDM results for the NICER 0.3--8~keV light curve (left panel), for the NuSTAR 3--10~keV light curve (middle panel) and for optical $V$-band light curve (right panel) during the 2023 outburst in GK Per.
Upper and lower panels represent $\Theta$-diagrams of our PDM analyses and phase-averaged profiles, respectively.
The cyan region in the upper panels represents the 90\% confidence interval.
}
\label{pdm}
\end{figure*}

We searched periodic signals from X-ray and optical light curves by using the phase dispersion minimization (PDM) method \citep{PDM}.
Before applying PDM, we subtracted the long-term trend of the light curve by locally weighted polynomial regression (LOWESS: \cite{LOWESS}).
We determined the length of the data for which the regression should be performed, and the smoother span ($f$) influencing the smoothness of each data point in this method.
In this work, a smoother span ranges between 0.1 and 0.2.
We detected the WD spin period of 351.32(8)~s from NICER 0.3--8~keV light curves and a 5699(2)-s period from AAVSO light curves (see the left and right panels of Figure \ref{pdm}).
The 1$\sigma$ error was computed via the method in \citet{fer89error}.
Although the latter period was not detected from NICER light curves even during the bright state of this outburst, $\sim$4000--6000-s QPOs were detected from X-ray light curves in previous outbursts \citep{mor99gkperQPO,vri05gkper,zem17gkper}.
No detection of this period from NICER light curves would be an observational bias since one observational visit by NICER is only $\sim$1000~s, and the time interval between two visits is at least $\sim$1.5 hrs long.

We have applied the same method for NuSTAR 3--10~keV light curves and detected a 351.32(6)-s period that is consistent with the WD spin period detected from NICER light curves (see the middle panel of Figure \ref{pdm}).
We did not detect any QPOs from NuSTAR light curves.
Since the NuSTAR data were taken at the end of the outburst, we could not investigate whether QPOs were observable in X-rays during this outburst or not.
The continuous observation by Tomo-e Gozen light curves lasted only for $\sim$15 min and was too short to search QPOs.
The WD spin signal was not detected at optical wavelengths. 
Other signals having $\sim$320-s and $\sim$380-s periods were reported during some time intervals in previous outbursts of GK Per at optical and X-ray wavelengths \citep{mor99gkperQPO,nog02gkper}.
However, we did not detect any signals having these periods.

\subsection{Energy dependence and time evolution of WD spin pulses} \label{sec:res-spin-pulse}

\begin{figure*}[htb]
\begin{minipage}{0.49\hsize}
\plotone{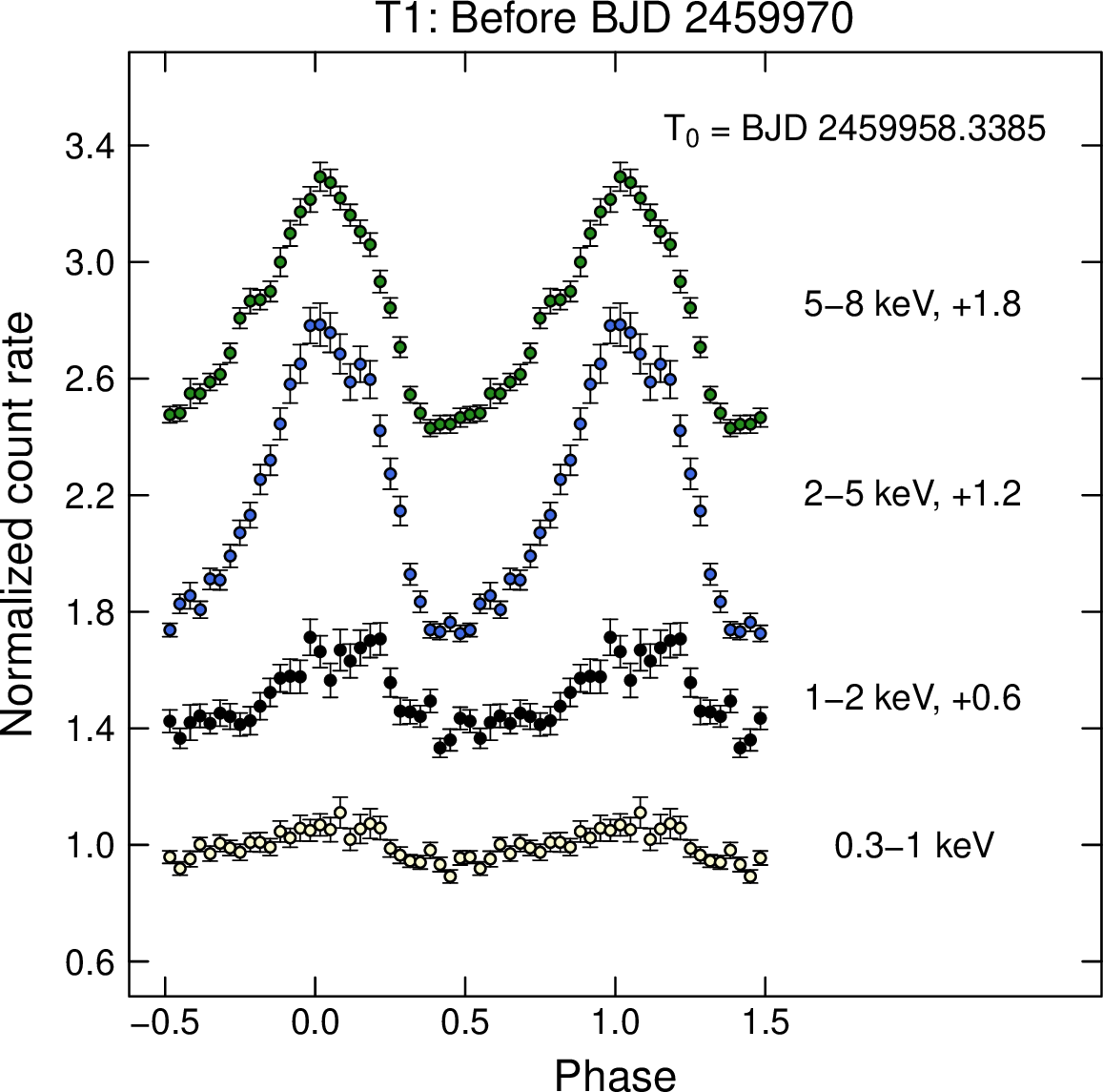}
\end{minipage}
\begin{minipage}{0.49\hsize}
\plotone{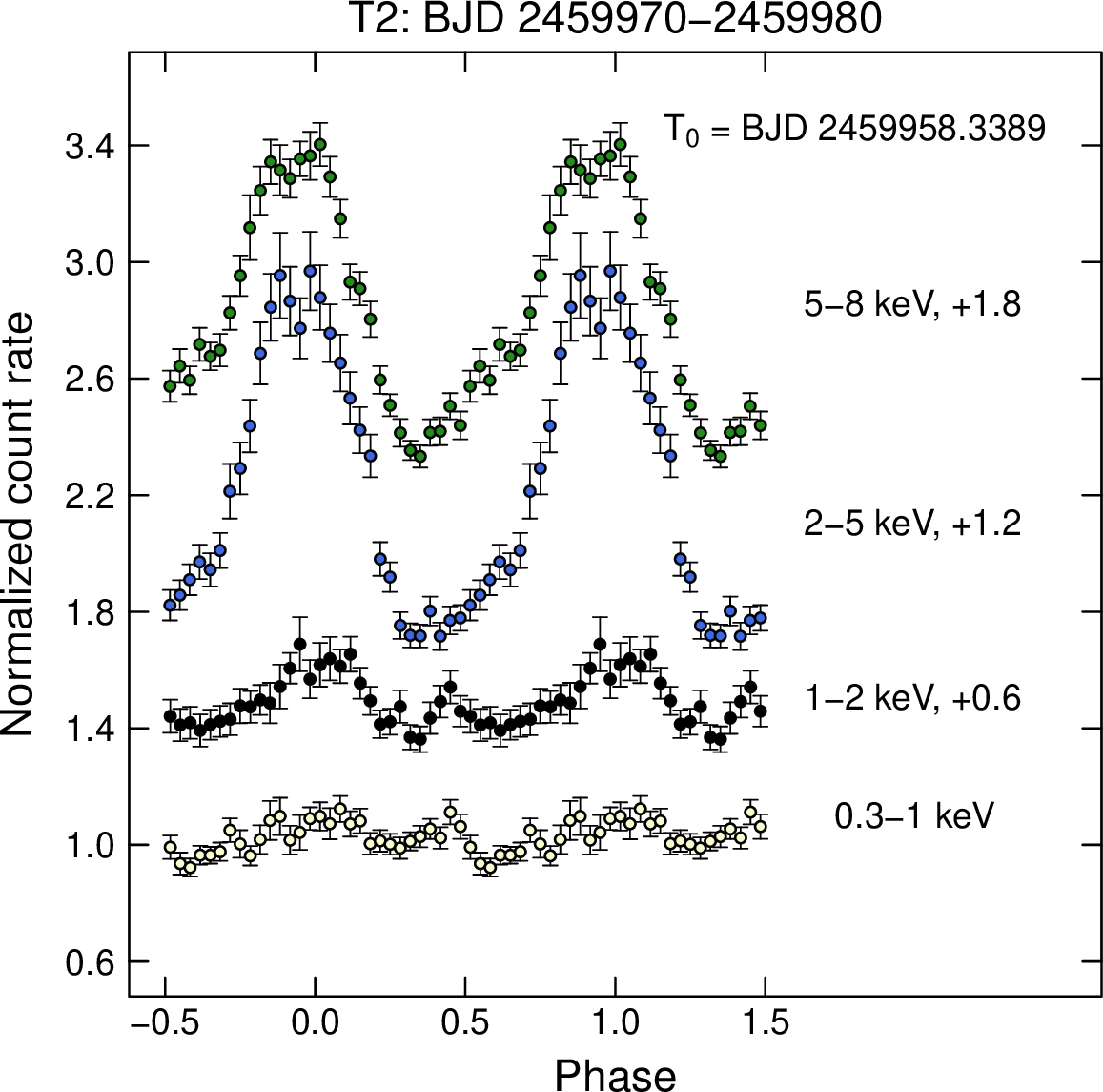}
\end{minipage}
\\
\begin{minipage}{0.49\hsize}
\plotone{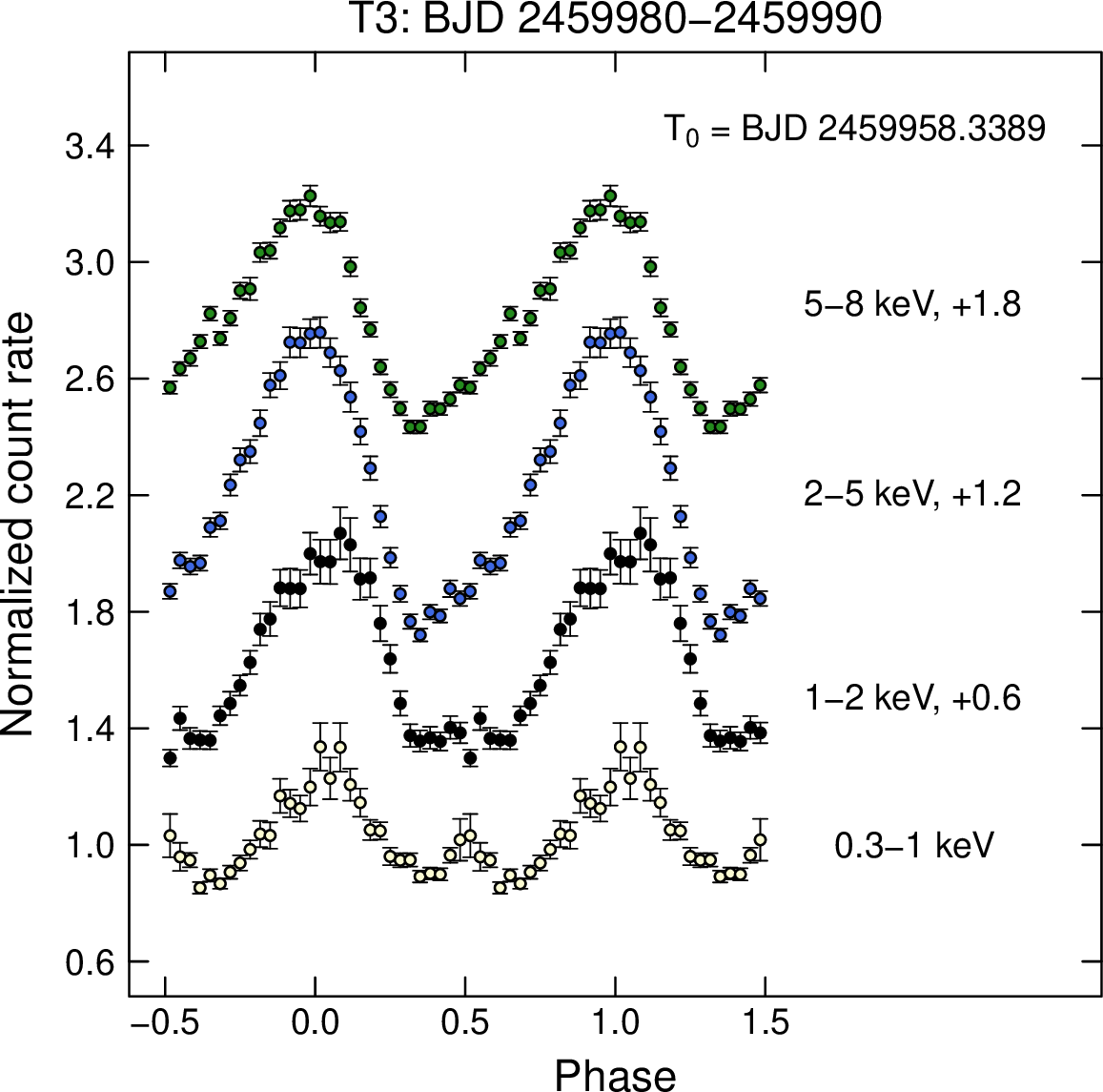}
\end{minipage}
\begin{minipage}{0.49\hsize}
\plotone{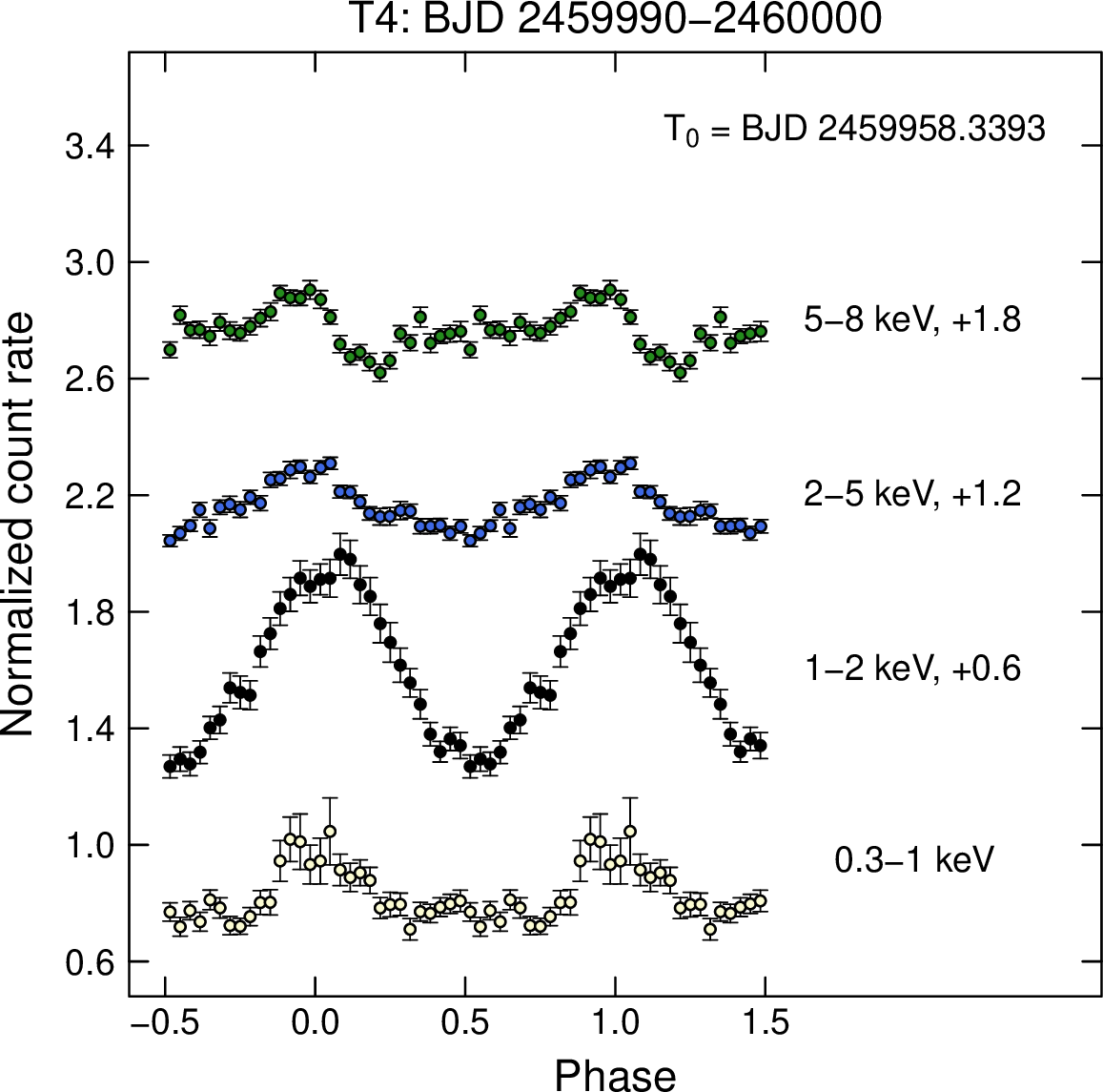}
\end{minipage}
\caption{
Phase-averaged profiles of WD spin pulses during the 2023 outburst in GK Per in T1, T2, T3, and T4 in four energy bands.
The $T_0$ value is given in each panel.
The 1--2~keV, 2--5~keV, and 5--8~keV profiles are shifted vertically for visibility.
These offset values are given in each panel.
}
\label{spin-profile-energy}
\end{figure*}

We divided the NICER data into four time zones, T1: before BJD 2459970, T2: BJD 2459970--2459980, T3: BJD 2459980--2459990, and T4: BJD 2459900--2460000, and made phase-averaged profiles of the WD spin pulse in four energy bands (0.3--1 keV, 1--2 keV, 2--5 keV, and 5--8 keV) in each time zone.
These time zones are denoted at the top panel of Figure \ref{overall}).
Here, we fixed the spin period at 351.32~s because we could not detect any variations of this period.
We defined the epoch of the light maximum of the phase-averaged profile in the 0.3--8~keV band as $T_0$.

Figure \ref{spin-profile-energy} shows the resultant pulse profiles.
The pulse amplitude at $<$1~keV was consistently lower than 20\%.
Although the pulse was dominant at more than 2 keV till T2, the pulse amplitude in the 1--2~keV band became higher after that.
The amplitude was highest around the outburst maximum, and the signal rapidly disappeared after T4.
The pulse profile was basically single-peaked and not symmetric to phase 0. 
Double-peaked humps occasionally appeared, e.g., the 0.3--1~keV phase-averaged profile in T3.
A flat-bottomed profile was observed in T3 in the 1--2~keV band.
The rising slope to the light maximum was less steeper than the fading slope to the light minimum.
The pulse peak in the 0.3--1~keV and 1--2~keV bands shifted to later phases with respect to those in the 2--5~keV and 5--8~keV bands.
The shift of $T_0$ was 0.2 phases from T1 to T4, which would come from the accuracy of our estimated spin period.

\begin{figure*}[htb]
\epsscale{0.5}
\plotone{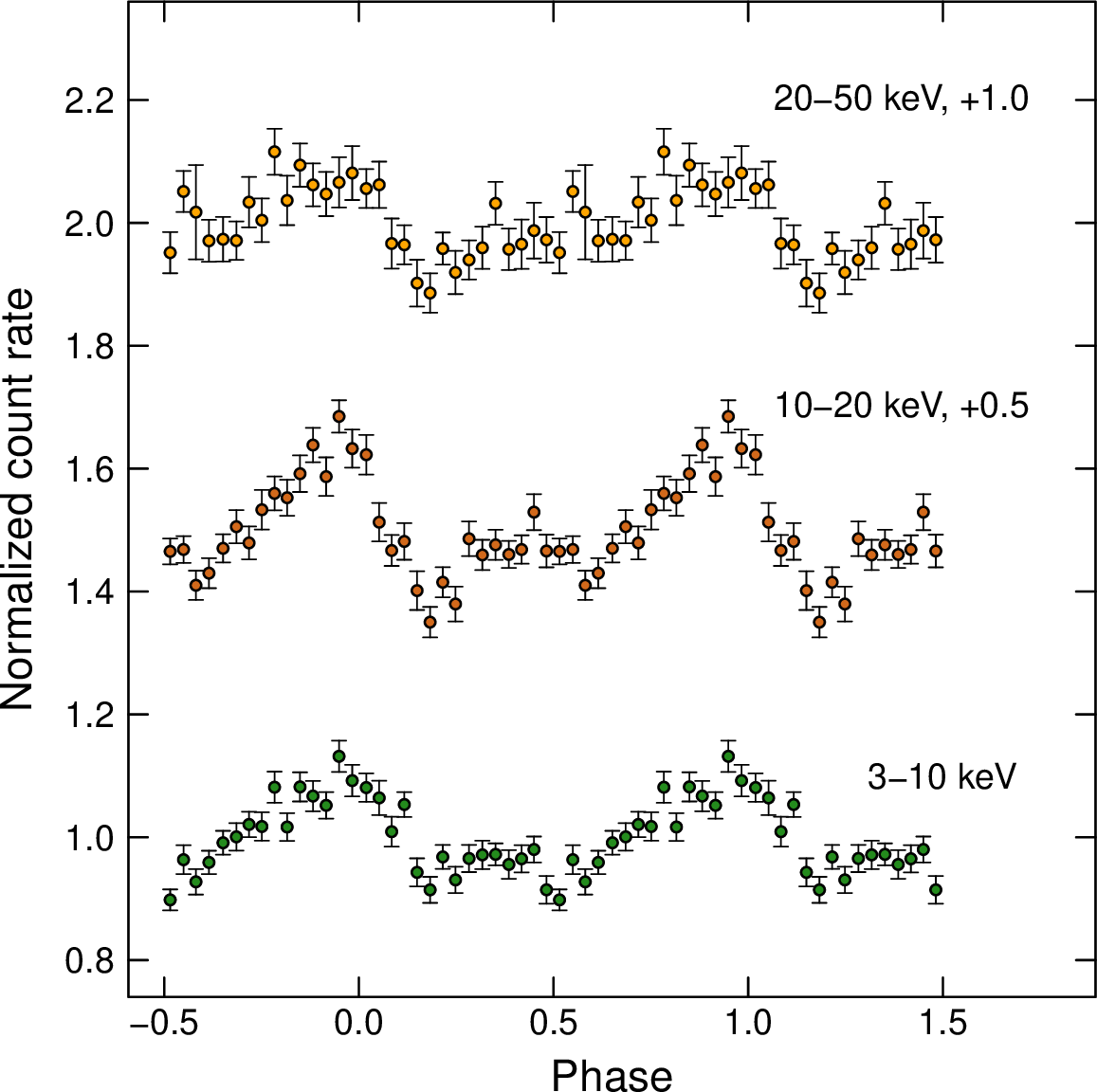}
\caption{
Phase-averaged profiles of WD spin pulses at the end of the 2023 outburst in GK Per in three energy bands.
The 10--20~keV and 20--50~keV profiles are shifted vertically for visibility.
These offset values are given in this plot.
}
\label{nustar-pulse}
\end{figure*}

We also investigated the energy dependence of pulse profiles in harder energy bands by using the NuSTAR data.
Figure \ref{nustar-pulse} shows the result.
The epoch is the same as that of T4, as determined by the NICER light curve.
The profile was double-peaked and the amplitude had almost no energy dependence.

\begin{figure*}[htb]
\epsscale{1.0}
\begin{minipage}{0.49\hsize}
\plotone{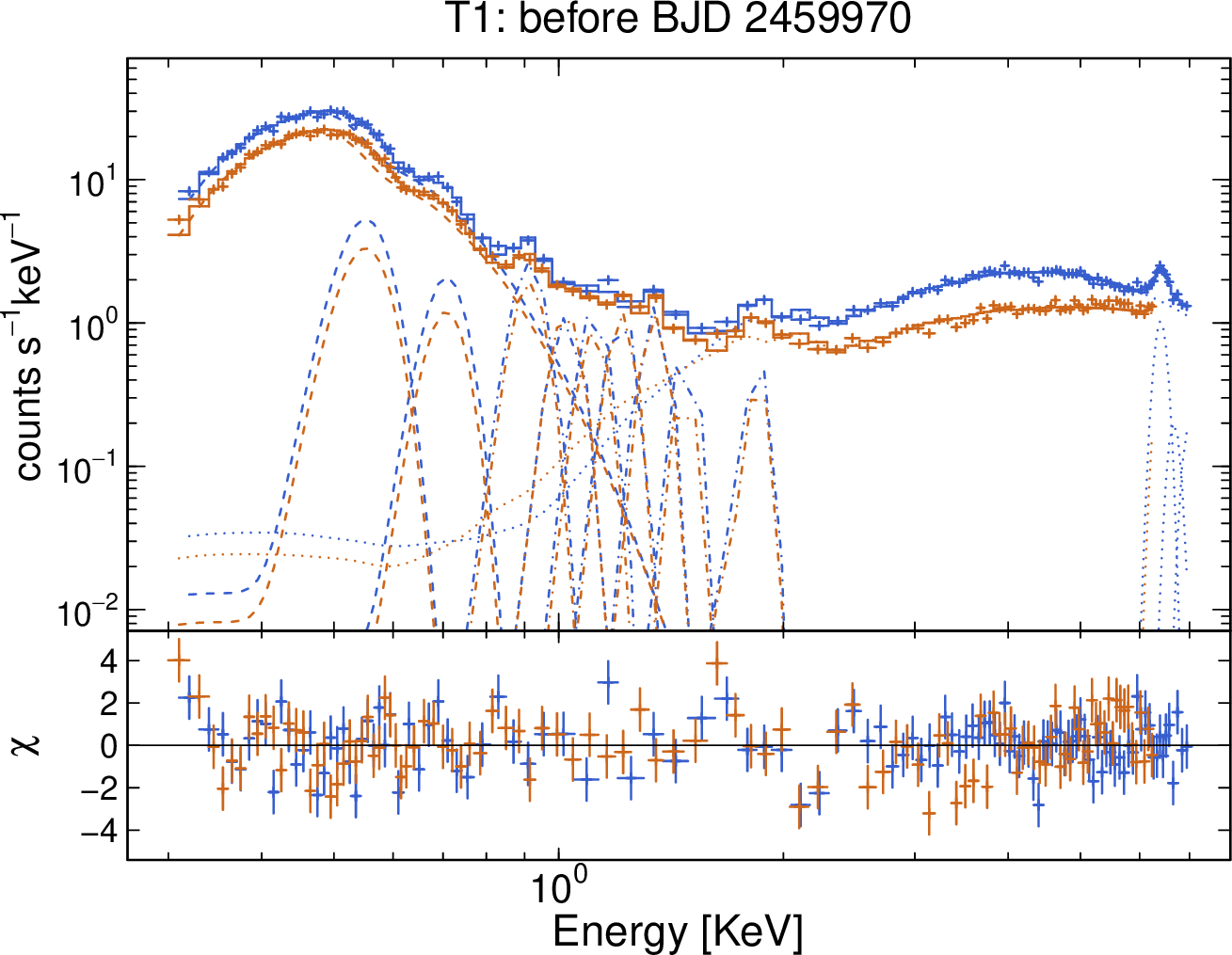}
\end{minipage}
\begin{minipage}{0.49\hsize}
\plotone{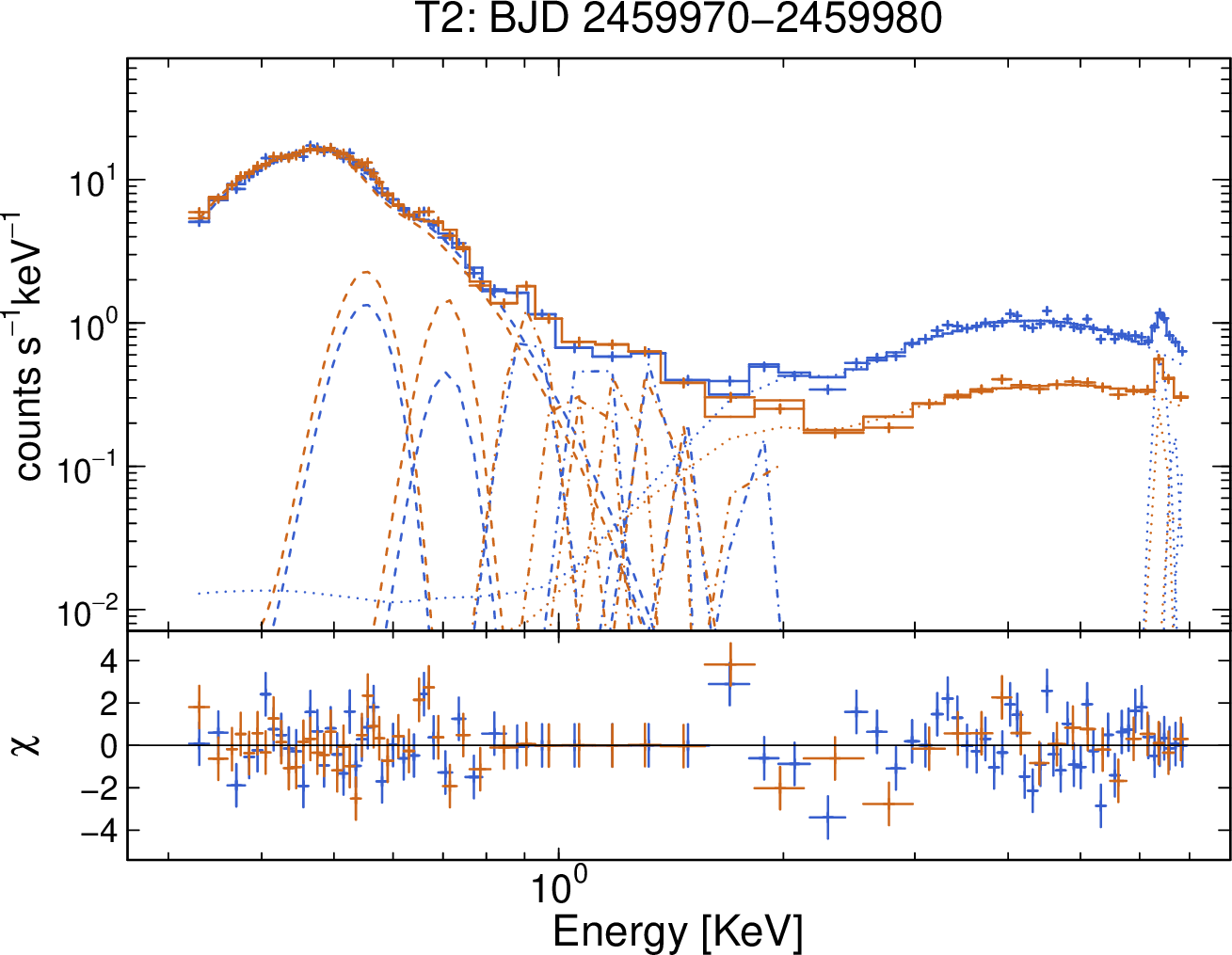}
\end{minipage}
\\
\epsscale{1.0}
\begin{minipage}{0.49\hsize}
\plotone{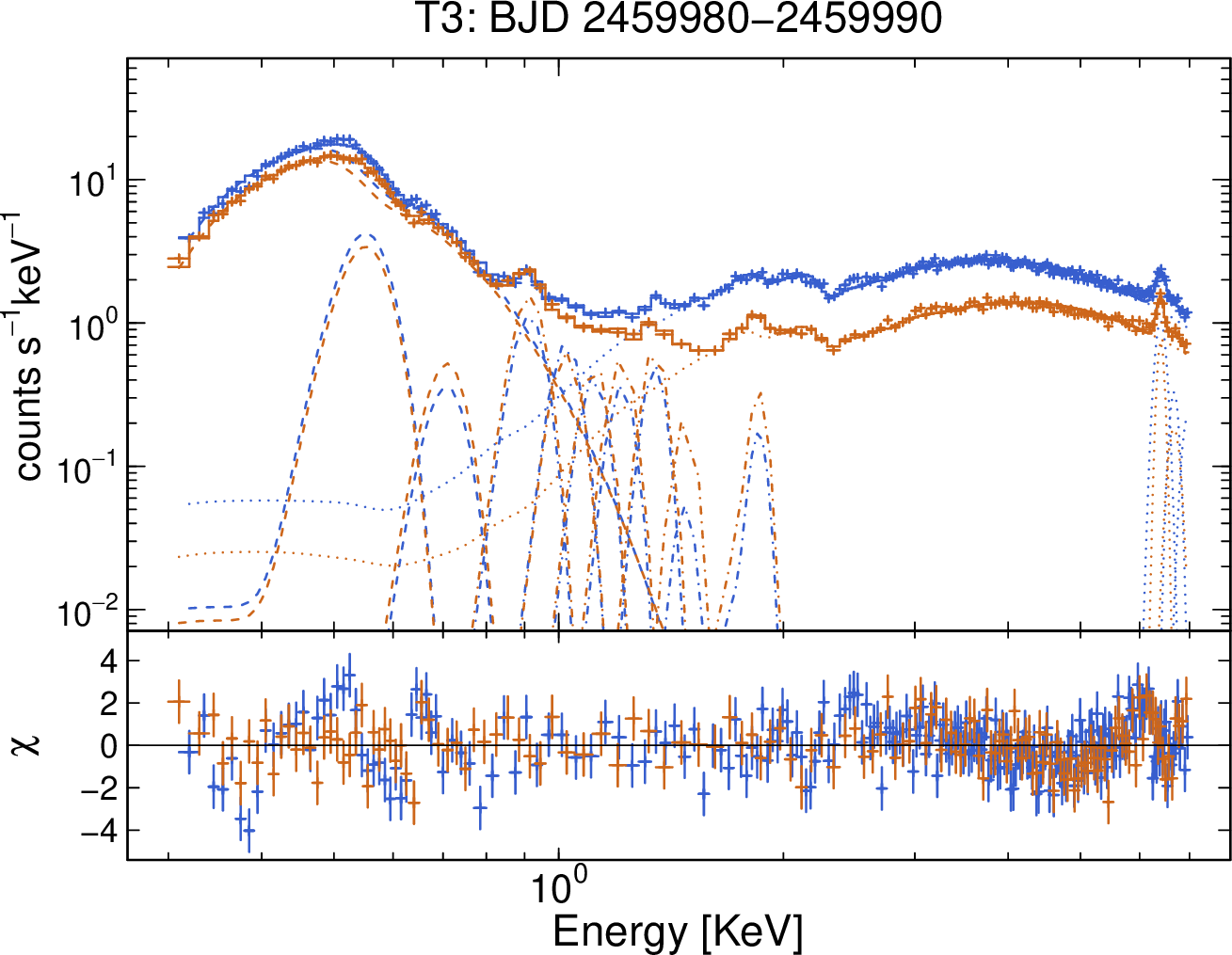}
\end{minipage}
\begin{minipage}{0.49\hsize}
\plotone{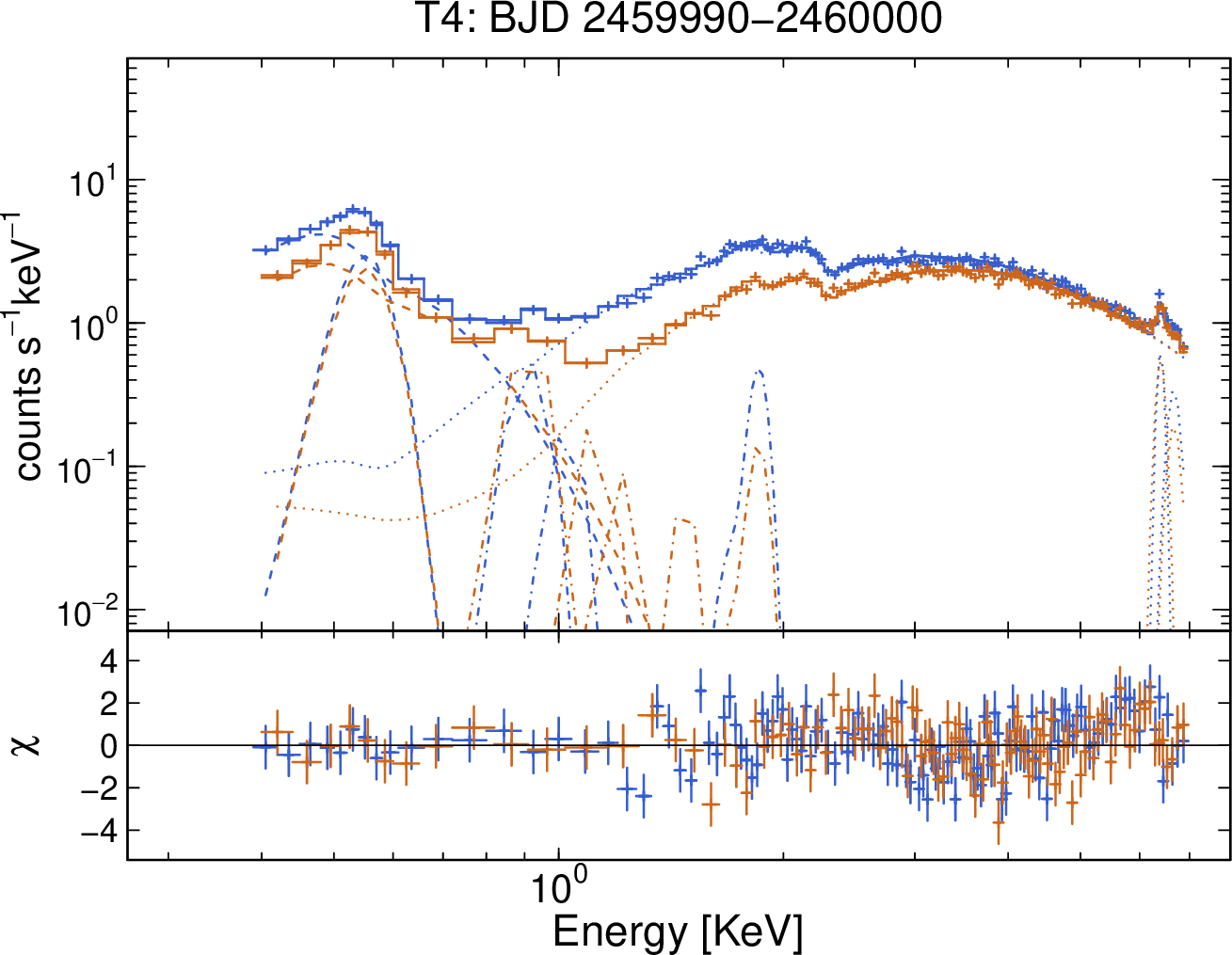}
\end{minipage}
\caption{
X-ray spectra at the on-pulse and off-pulse phases of GK Per during the 2023 outburst, which are overlaid with the best-fit spectral model of \texttt{Tbabs*(gaussian + gaussian + bbody + gaussian + gaussian + gaussian + gaussian + gaussian + gaussian + gaussian + pwab*(reflect*powerlaw + gaussian + gaussian + gaussian))}. 
The blue and orange colors denote the data and models during the on-pulse and off-pulse phases, respectively.
The cross represents the averaged NICER spectrum. 
The dot-dashed line denotes the best-fit models of neon, iron, magnesium, and silicon emission lines.
The dashed line represents the best-fit model of blackbody emission with oxygen and iron lines.
The dot line represents the best-fit model components of hot plasma plus three iron lines.
The solid line shows the total best-fit model emission. 
}
\label{spec-10d-spin-on-off}
\end{figure*}

We next tried the modeling of on-pulse and off-pulse spectra in T1--T4.
Here, on-pulse and off-pulse spectra indicate spectra for the spin pulse phase $-$0.1--0.1 and the spin pulse phase 0.4--0.6, respectively.
The applied model is the same as that used in section \ref{sec:evolution-xray-spec}.
The spectra and the best-fit models are shown in Figure \ref{spec-10d-spin-on-off}.
The PL flux drastically changed between the on-pulse and off-pulse phases, while the the BB and line fluxes did not change very much.

\subsection{Correlation between X-ray and optical fast variability} \label{sec:fast-variability}

\begin{figure*}[htb]
\epsscale{1.0}
\begin{minipage}{0.49\hsize}
\plotone{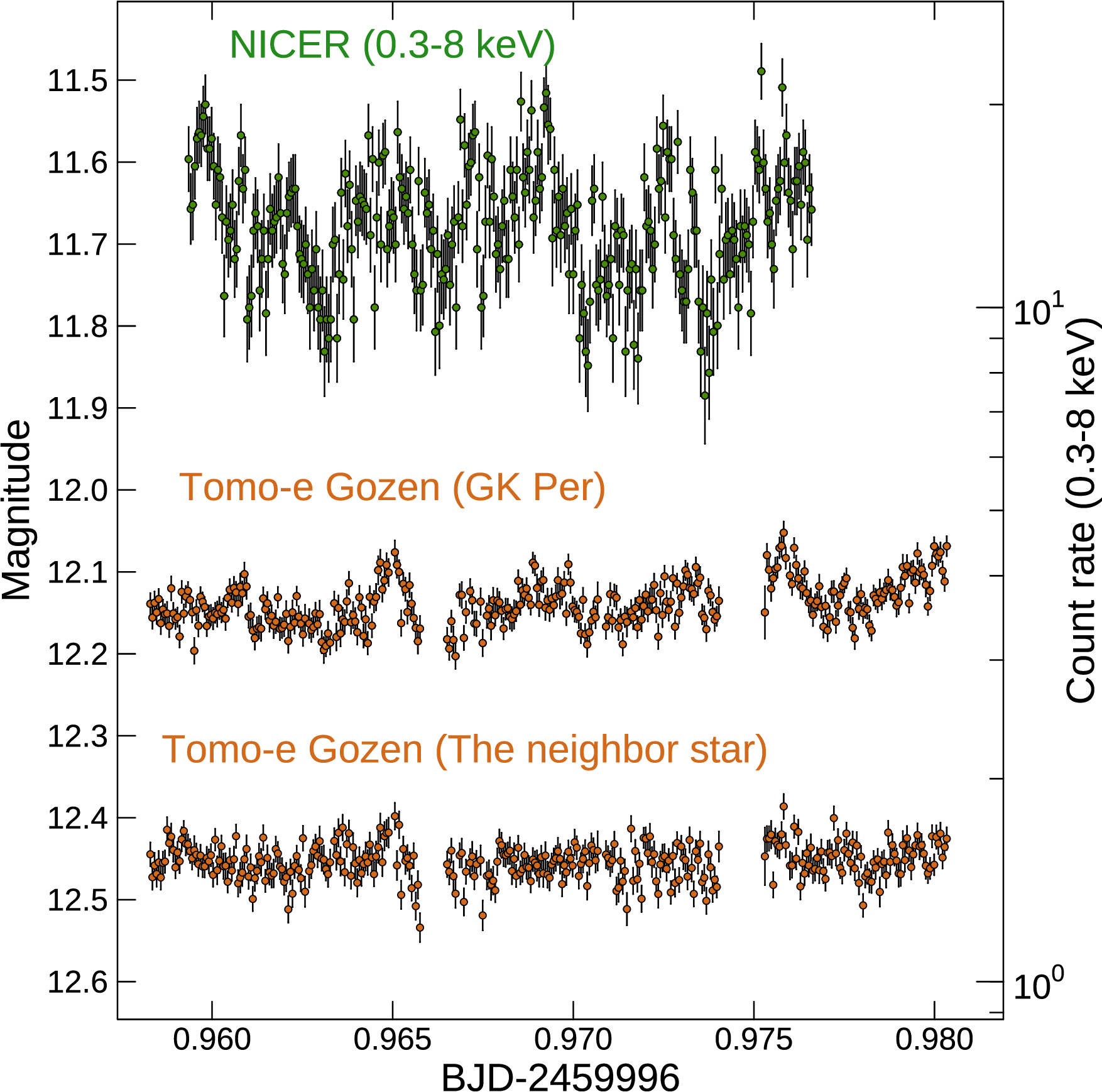}
\end{minipage}
\begin{minipage}{0.49\hsize}
\plotone{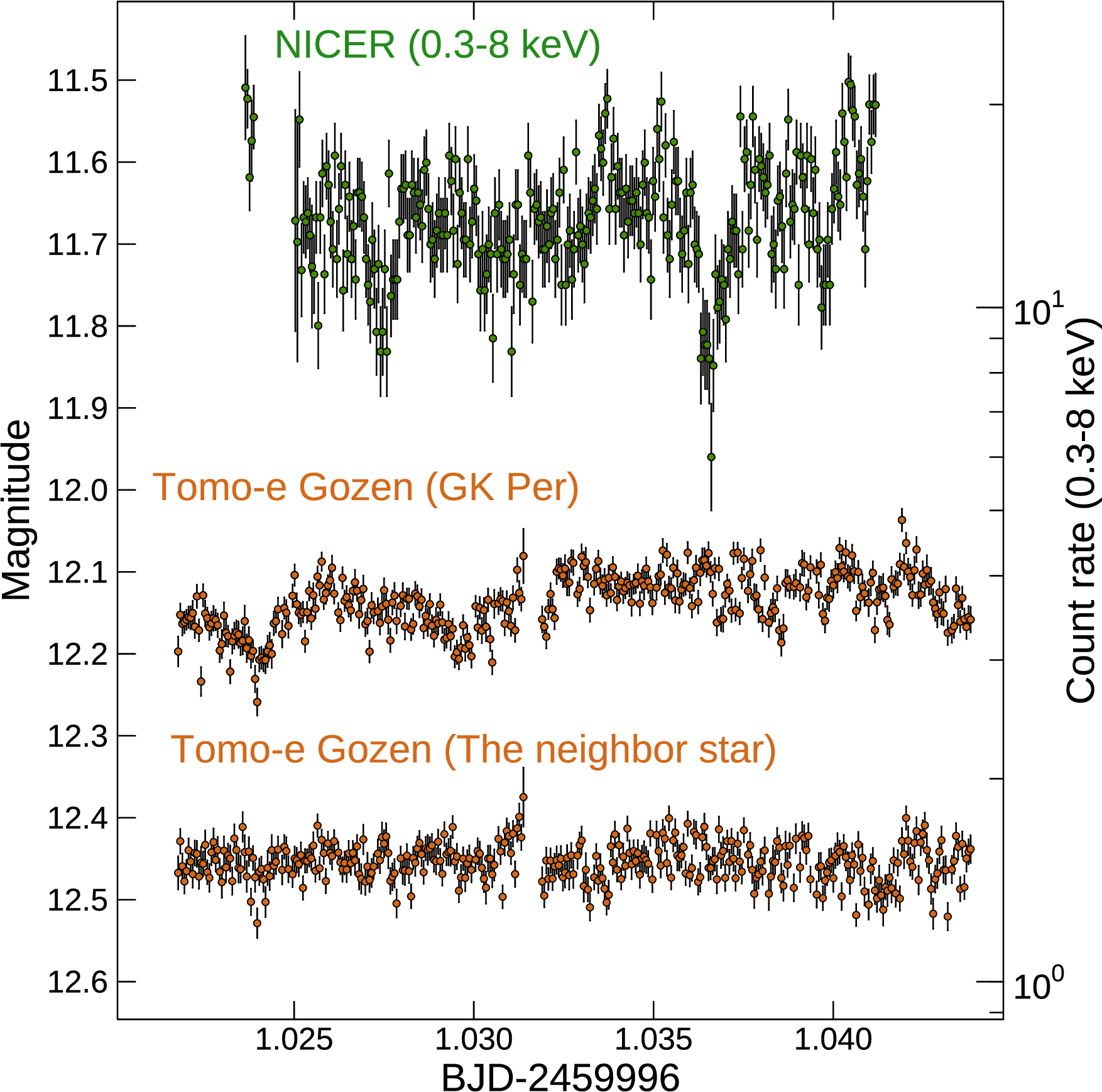}
\end{minipage}
\caption{
X-ray and optical simultaneously observed light curves taken by NICER and Tomo-e Gozen on 2023 February 21.
The green and orange dots represent the NICER and Tomo-e Gozen light curves, respectively.
We also plot the Tomo-e Gozen light curve of the neighbor star USNO A2.0 1275-02381887.
}
\label{fast-variability}
\end{figure*}

High-speed optical photometry of GK Per with Tomo-e Gozen was performed on four nights, and totally two visits were completely overlapped with the NICER observations.
The simultaneously observed X-ray and optical light curves are displayed in Figure \ref{fast-variability}.
Here, the optical light curve was binned in 5-s bins as the NICER light curve was.
The amplitude of optical variations was much smaller than that of X-ray ones in these two windows.
The variations of the optical light curve of GK Per was larger than that of the neighbor star.
However, atmospheric fluctuations would affect the observed light curve of GK Per since the light curve of the neighbor star also fluctuated by 0.04~mag within 2$\sigma$.
Although the gradual increase of the flux with timescales of $\sim$0.01~d would be intrinsic variations of GK Per, the steep drop of the flux around BJD 2459996.966 and the sharp increase of the flux around BJD 2459997.031 would be a kind of atmospheric fluctuations.
The intrinsic variations are likely flickering well-observed in this system \citep{bru21flikering}.

\begin{figure*}[htb]
\epsscale{1.0}
\begin{minipage}{0.49\hsize}
\plotone{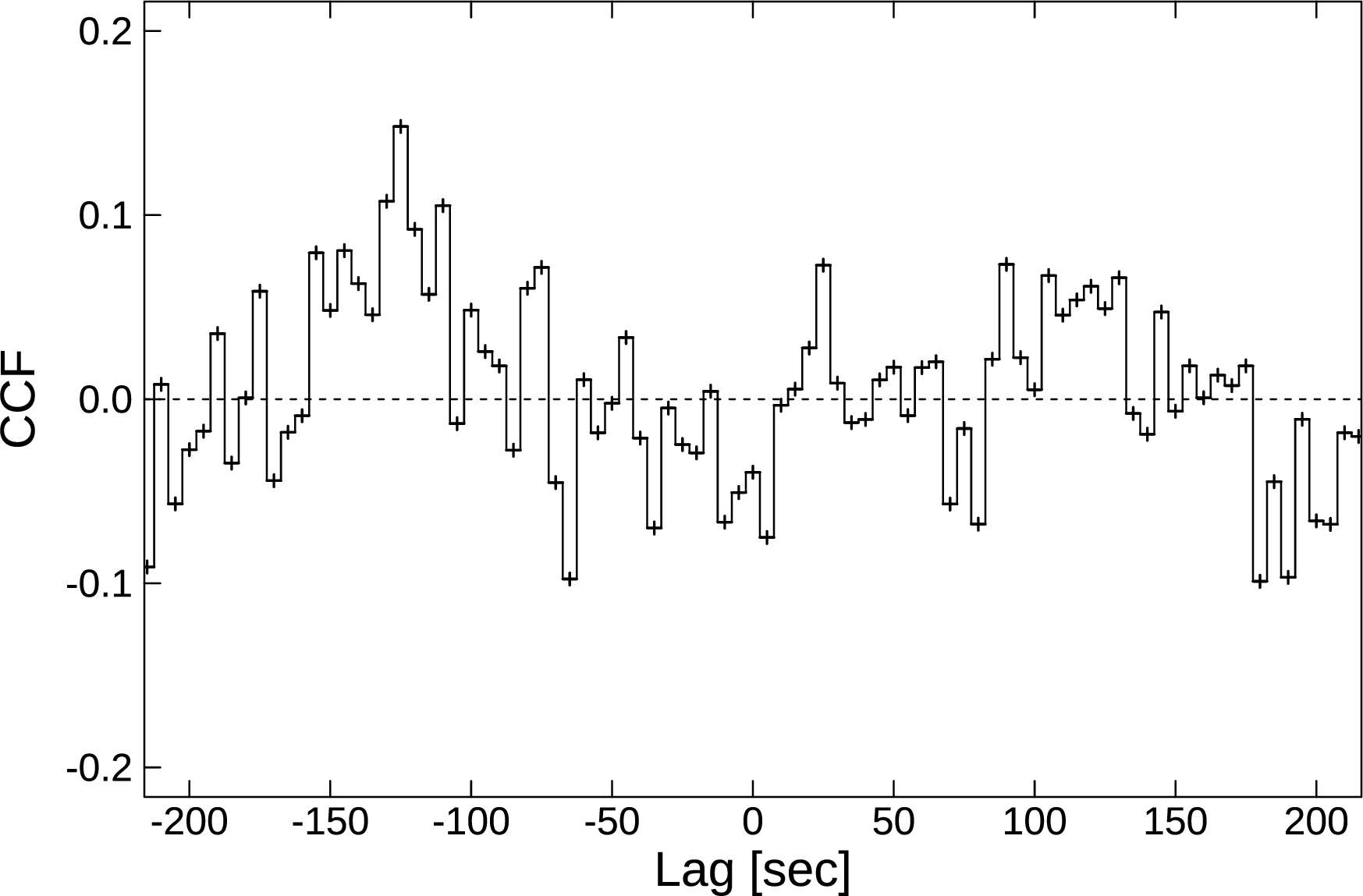}
\end{minipage}
\begin{minipage}{0.49\hsize}
\plotone{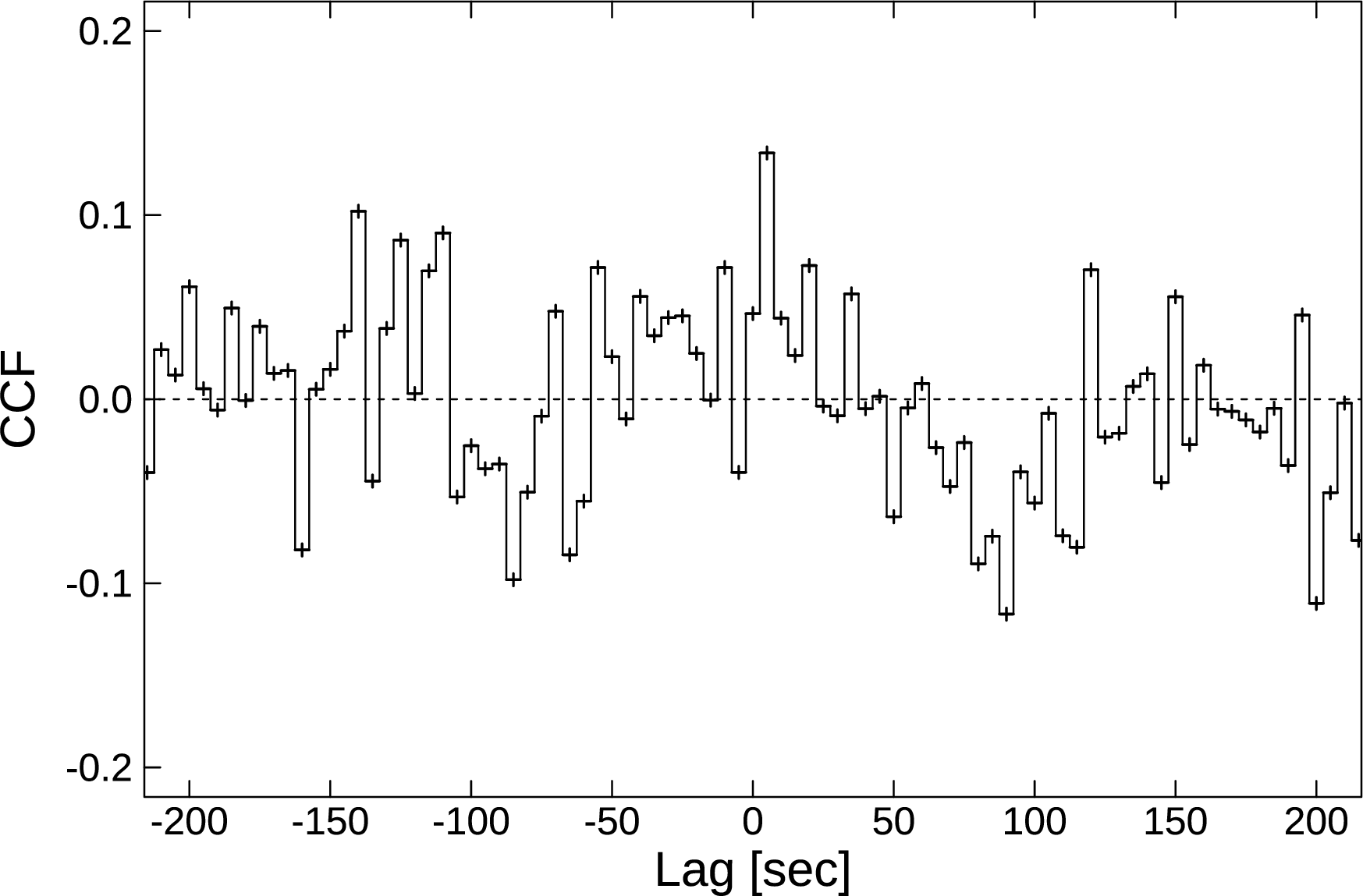}
\end{minipage}
\caption{
CCFs between the NICER and Tomo-e Gozen light curves of GK Per on 2023 February 21.
The positive lag indicates the delay of optical variations to X-ray ones.
The data in the left and right panels correspond to the light variations in the left and right panels of Figure \ref{fast-variability}.
}
\label{ccf}
\end{figure*}

We investigated the correlation between the X-ray and optical modulations by the cross-correlation function (CCF) analysis.
The calculation method for CCF is described in e.g.,~\citet{pat85CVXrayemission1,gan10gx339}.
The optical light curves are resampled to a 5~s time bin which is the same as the X-ray one.
Before CCF analyses, we subtracted the long-term light curve trend from these light curves by using LOWESS \citep{LOWESS} as done in section 3.4.
The smoother span ranged between 0.2 and 1.0.
The resultant CCFs are given in Figure \ref{ccf}.
The positive lag represents the lag of optical variations to X-ray variations.
No significant correlation was detected.

\section{Discussion}

\subsection{Origins of X-ray emission sources} \label{sec:discuss-xray-spec}

We performed the modeling of X-ray spectra in sections \ref{sec:sed} and \ref{sec:evolution-xray-spec}.
The main emission components are (1) BB emission dominant at $<$1 keV, (2) several emission lines of neon, magnesium, iron, and silicon, which are dominant in the 1--2~keV range, and (3) multi-temperature bremsstrahlung emission dominant at $>$2 keV. 
The component (3) is regarded as the emission from the high-temperature and optically-thin AC. 
The component (1) is interpreted as the emission from a part of the WD surface reprocessed and heated by hard X-ray photons from the AC.

The origin of the component (2) is under debate \citep{zem17gkper,pei24gkper}.
The absorption was ineffective for the source (2) at the off-pulse phase during the 2023 outburst (see section \ref{sec:res-spin-pulse}), which suggests that the emission source was located outside of the curtain.
\citet{zem17gkper} showed no excess of nitrogen abundance, which implies that the emission source is unlikely a nova shell irradiated by hard X-rays from the AC.
We consider that the possible origin is photoionization at the time-varying low-temperature gas of the inner region of the accretion disk and/or the disk wind since the line emission became much weaker at the fading stage of this outburst despite the little change in the hard X-ray emission from the AC (see Figures \ref{spec-5d-04-08}).
For instance, \citet{muk03CVXray} found that the high-resolution spectrum in GK Per resembled the expectation for photoionized plasma.
However, our results cannot settle the issue of whether collisional ionisation or photoionisation reproduces the spectrum in 1--2~keV (see also the texts in section 3.3).

\subsection{Interpretations of X-ray light curves} \label{sec:discuss-xray-lcs}

On the basis of interpretations in section \ref{sec:discuss-xray-spec}, we can regard that Figure \ref{modelflux} shows the time evolution of X-ray emission from the three main components of the WD surface, the line emitter, and the AC.
We interpret below what the three X-ray light curves in the bottom panel of Figure \ref{overall} represent.
First of all, we consider the time evolution of the AC flux with increasing mass accretion rates onto the WD, denoted by $\dot{M}_{\rm acc}$.
If $\dot{M}_{\rm acc}$ becomes higher, the column density of the curtain becomes higher.
More X-ray photons from the AC are absorbed.
The absorption is effective not only for the 2--5 keV band but also for the 5--8 keV band.
The 2--5~keV and 5--8~keV model fluxes of the PL component were lower if the system was brighter at optical wavelengths (see the top and the fourth panels of Figure \ref{modelflux}).
The 2--8~keV light curve in the bottom panel of Figure \ref{overall} reflected the AC flux absorbed by the curtain.
The flux dropped around the outburst maximum since the column density of the curtain increased with the increase in $\dot{M}_{\rm acc}$.

To interpret the unabsorbed PL flux in the fifth panel of Figure \ref{modelflux}, the increase in soft X-ray photons accompanied by the shrinkage in the inner disk edge should be taken into account.
If $\dot{M}_{\rm acc}$ becomes higher, the ram pressure is higher, and the radius of the inner disk edge ($r_{\rm in}$) becomes smaller.
Then, the shock temperature ($T_{\rm sh}$) of the AC decreases since the free-fall velocity of the gas decreases (see, e.g., equation (5) in \citet{wad18gkper}).
As a result, the X-ray spectrum would become softer.
The inferred X-ray luminosity would decrease if the AC becomes optically thick.
This phenomenon could explain why the intrinsic X-ray flux from the AC decreased if the system became brighter in optical.

Next, we consider the evolution of the BB flux originating from the irradiated WD surface, as imprinted on the 0.3--1~keV light curve in the bottom panel of Figure \ref{overall}.
The evolution of the BB flux should be related to that of the AC via the irradiation process.
The BB flux decreased around the outburst maximum when the unabsorbed flux of the PL component of $>$1~keV decreased.
This would be because hard X-ray photons from the AC decreased.
The reason for the smaller decrease in the BB flux than that in the AC flux is probably because a part of the AC became optically-thick and increased the BB flux.

The BB flux rapidly decreased at the later part of the fading stage, after S6, though the flux level of the PL component at that time was the same level before the outburst maximum.
If $\dot{M}_{\rm acc}$ decreases and $r_{\rm in}$ increases, the reflection effect becomes stronger since the spectrum becomes harder with the increase in $T_{\rm sh}$.
The spectral hardening from the fading stage of the outburst to the quiescence was reported in previous works \citep{wad18gkper,pei24gkper}.
The electron scattering becomes more dominant than the photoelectric absorption at higher energies, $\gtrsim$10~keV \citep[see Figure 2 in][]{hay18reflection} and the irradiation effect becomes lower, which results in a drop of the BB flux.

Finally, we consider the evolution of the line flux.
The line flux seems to have traced the absorbed flux of the PL component at $>$2~keV (see the black and white dots in the third panel and the blue and green dots in the fourth panel of Figure \ref{modelflux}).
However, the rapid decrease of the line flux after S6 is not straightforward, as discussed above for the BB flux.
The time variation of the photo-ionized gas may naturally explain this phenomenon.
The inner disk edge and the disk wind would become more distant from the AC if $\dot{M}_{\rm acc}$ decreased and $r_{\rm in}$ increased.
In this case, photoionization becomes weaker even if the X-ray flux from the AC does not change.
The 1--2~keV light curve in the bottom panel of Figure \ref{overall} included both the emission from the AC and that from the line emitter.
The emission from the AC became dominant after the middle of the fading stage (see the bottom panel of Figure \ref{modelflux}).
The flux slightly increased at the end of this outburst because the column density of the curtain became lower with the decrease in $\dot{M}_{\rm acc}$.

\subsection{Can the evolution of WD spin pulses be explained by the accretion-curtain model ?} \label{sec:discuss-spin-pulse}

As shown by the modeling of X-ray spectra, the BB and line fluxes did not change very much between the on-pulse and off-pulse phases (see Figure \ref{spec-10d-spin-on-off}).
The main source of the WD spin pulse is the change in the AC flux to the line of sight.
The column density of the absorber was much higher in the off-pulse phase than in the on-pulse phase, which suggests that the absorption of the AC created the pulse signal during the 2023 outburst in GK Per.

Since the AC emission was dominant at $>$2 keV in T1 and T2, the amplitude of spin pulses of $<$2 keV was low.
After the outburst maximum, the AC flux overwhelmed the line flux even in the 1--2~keV band (see the bottom panel of Figure \ref{modelflux}), and the pulse amplitude in the 1--2~keV band increased in T3 (see Figure \ref{spin-profile-energy}).
The brighter the system was at optical wavelengths, the higher the pulse amplitude of $>$2~keV was, which is consistent with that more X-ray photons are absorbed with increasing $\dot{M}_{\rm acc}$.
The growing pulse amplitude with the increasing mass accretion rate during the 2018 outburst in this system was reported by \citet{pei24gkper}.
The amplitude was higher in the 2--5~keV band than in the 5--8~keV band (see Figure \ref{spin-profile-energy}).
This energy dependence was weaker when the system was brighter (see also Figure \ref{spec-10d-spin-on-off}), which can be explained by the change in the column density of the curtain. 

If we can observe both of the two magnetic poles of the WD, the pulse profile should be double-peaked.
However, the pulse profile is basically single-peaked (see Figure \ref{spin-profile-energy}).
This would be because the magnetic pole below the disk equatorial plane is hidden by the extended inner part of the disk.
This idea was proposed by \citet{hel04gkper}.
The profile at the end of the 2023 outburst became double-peaked (see Figure \ref{nustar-pulse}), which would indicate the expansion of the inner disk edge due to the decrease in $\dot{M}_{\rm acc}$.
On the other hand, a small hump around the phase 0.5 was observed in T3 in the 0.3--1~keV energy band.
The irradiated WD surface was wider than the AC, and it below the disk equatorial plane might be observable.
Although the BB emission was not sensitive to the absorption by the curtain, the change in the surface area visible to the observer can change between the on-pulse and off-pulse phases.

\begin{figure*}[htb]
\epsscale{0.5}
\plotone{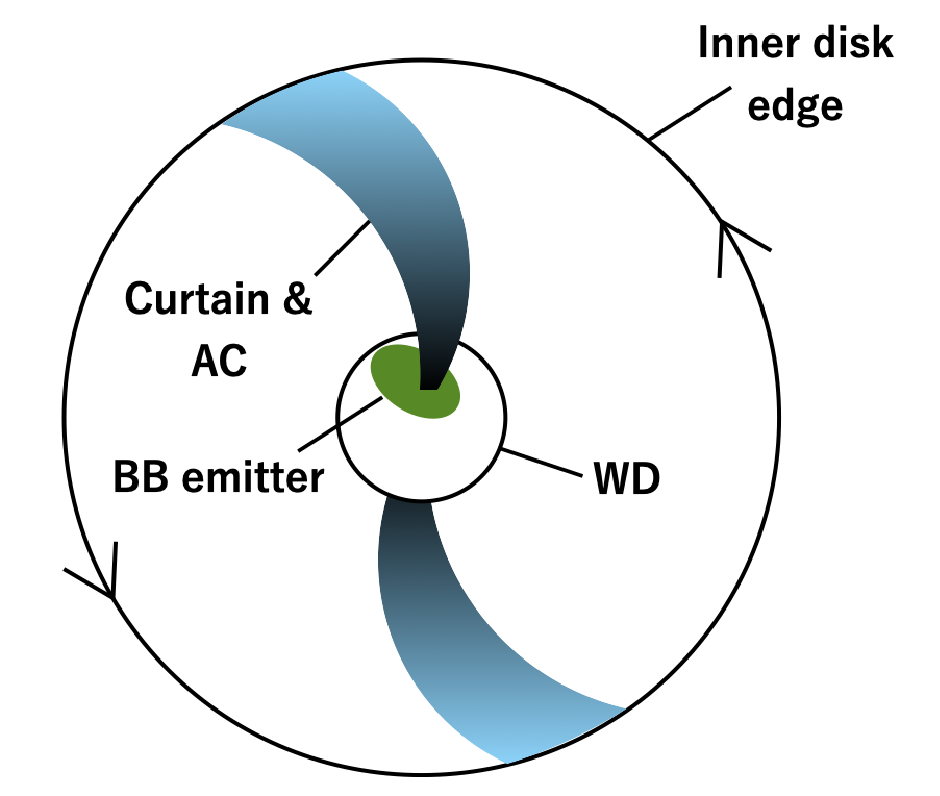}
\caption{
Schematic figure of the face-on view of the twisted geometry of the AC and the curtain during the 2023 outburst in GK Per.
The inner disk edge rotates counterclockwise.
The irradiated WD surface is asymmetric to the center of the AC.
}
\label{twisted-geometry}
\end{figure*}

The pulse profile was not symmetric to the peak phase, and the peak phase in the lower energy bands shifted towards later phases to that in the higher energy band (see Figure \ref{spin-profile-energy}).
These observational features could be explained by the twisted geometry of the AC and curtain (see the schematic figure in Figure \ref{twisted-geometry}).
If the inner disk edge is inside of the co-rotation radius in outbursts, the edge rotates faster than the WD.
The magnetic field line from the WD surface is tethered to the disk inner edge and the disk gas accretes onto the WD along the field line.
The field line is dragged towards the rotational direction.
The WD surface irradiated by the AC would be also asymmetric to the center of the AC base.

The above-discussed observational features can be basically explained by the absorption by the curtain.
However, we need additional factors to interpret almost no energy dependence of the high-energy pulse profile at the end of the outburst (see Figure \ref{nustar-pulse}).
One possible source of this weak energy dependence is electron scattering.
According to Figure 2 in \citet{hay18reflection}, the mass attenuation coefficient of the incoherent scattering overwhelms that of the photoelectric absorption around 10 keV.
If the scattering is effective, we would see a high-energy pulse.
Another possible source is the change in the visible area of the AC to the observer between the on-pulse and off-pulse phases, which was pointed out by \citet{zem17gkper} as the geometrical effect.
If $\dot{M}_{\rm acc}$ decreases, the ram pressure decreases, and the scale height of the AC increases \citep{hya14exhyav1223sgr}.
The visible area of the AC to the observed is maximized at the on-pulse phase.
Since the column density of the curtain was low at the end of the outburst, this kind of geometrical effect might generate the pulse signal.

\subsection{Possible origins of optical QPOs} \label{sec:discuss-qpo}

We detected $\sim$5700-s QPOs from the optical light curve (see section \ref{sec:period}).
\citet{war02DNOproc} proposed that QPO signals are triggered by the disk oscillations, which make a vertically thick structure at a part of the disk.
The bump on the disk surface is irradiated by high-energy photons from the WD and the inner part of the disk, and the visible area of the irradiated surface of the bumpy structure changes with the Keplerian period at the radius of that structure.
For instance, if the gas transferred from the secondary star overflows the outer disk edge, the overflowing gas impacts around $r_{\rm min}$ given by $r_{\rm min} = 0.0488 q^{-0.464} a$, where $a$ is the binary separation \citep{war95book,kun01streamdiskoverflow}.
This process can generate a vertically thick structure at a part of azimuthal angle around $r_{\rm min}$ \citep{lub89overflow}.
In GK Per, $r_{\rm min}$ is 3.9$\times$10$^{10}$~cm.
The radius at which the Keplerian period is $\sim$5700~s is $\sim$5$\times$10$^{10}$~cm and close to $r_{\rm min}$.
The optical QPO might be caused by the irradiation at the vertically thick structure around this radius.
In this case, the QPO period would not change throughout the outburst.
We were not able to detect the time evolution of optical QPO periods due to our limited data.
More frequent observations are necessary in the future.

Here, we note that the irradiation by X-ray photons from the AC could not reproduce optical QPOs.
X-ray photons harder than $\sim$1~keV can irradiate the disk gas \citep{all59UVspec,cru74softXrayEUV}.
The X-ray luminosity in the 1--50~keV band was $\sim$4$\times$10$^{34}$~erg~s$^{-1}$ even at the brightest stage.
This was more than two orders of magnitudes lower than the intrinsic luminosity of the disk during the outburst in GK Per.
In fact, we did not find any strong correlations between the optical and X-ray rapid variations at the end of the outburst (see section \ref{sec:fast-variability}).
The main source of the irradiation would be ultraviolet photons from the WD surface and the inner part of the disk.

\subsection{Mechanism of outbursts in GK Per} \label{sec:outburst-mechanism}

Although the outburst in GK Per has some peculiar characteristics, it can be explained by the disk-instability model.
The $B-V$ color did not become redder at the onset of the 2023 outburst, which implies that the accretion disk did not drastically expand at the onset of the outburst.
The slow-rise optical light curve and slowly hardening PL component are prominent characteristics of inside-out outbursts, and suggest that the heating wave slowly propagated over the disk and that the temperature and the accretion rate of the disk slowly increased (see also Figures \ref{overall-optical} and \ref{modelflux}).
As discussed in \citet{kim18j1621}, inside-out outbursts tend to be triggered in long-period systems with low mass-transfer rates like GK Per.

Although the simple inside-out outburst shows a linear and slow brightness increase, the rising part of the 2023 outburst in GK Per seems to be complex, and at least two phases would exist.
This is interpreted as the representation of the stagnation phase.
The thermal equilibrium curve at the outer disk is $\xi$-shaped rather than $S$-shaped \citep{min83DNDI,can93DI}.
In this case, the outer disk does not always directly jump to the hot state and temporarily stays at the nonthermal equilibrium intermediate temperature state during the rising to the outburst maximum as demonstrated by \citet{kim92gkper} via numerical simulations.

\section{Summary}

We observed the 2023 outburst of GK Per by NICER, NuSTAR, and Tomo-e Gozen. 
Our major findings and their interpretations are as follows: 
\begin{itemize}

	\item The optical light curve was consistent with those in past outbursts of this object. We found a steeper rise followed by a slower rise to the outburst maximum. This would be because the outer disk did not directly jump to the hot state. The $B-V$ color evolution implies that the outburst was triggered at the inner part of the disk (see sections \ref{sec:overall} and \ref{sec:outburst-mechanism}).

	\item The broadband X-ray spectrum at the end of the outburst was mainly composed of two emission sources. One is a blackbody component with a temperature of $\sim$83~eV and the other one is a multi-temperature and optically-thin plasma with the maximum temperature of $\sim$54~keV. The former and the latter represent the emission from the WD surface and the AC, respectively (see sections \ref{sec:sed} and \ref{sec:discuss-xray-spec}). 

	\item The emission from the neon, magnesium, iron, and silicon lines were not negligible before the later part of the fading stage during the outburst. 
	These lines were not absorbed in the off-pulse phase, which suggests that they originate from the gas outside of the curtain. 


	\item The WD spin pulse with a 351.3-s period was observed only in X-rays. 
	The absorption of the AC was much larger at the off-pulse phase than that at the on-pulse phase. These observational features suggest that the spin pulse was reproduced by the absorption of X-ray photons from the AC, which is consistent with the accretion-curtain model (see sections \ref{sec:period}, \ref{sec:res-spin-pulse}, and \ref{sec:discuss-spin-pulse}).

	\item The evolution of the pulse amplitude at $>$2~keV was consistent with the change in the absorption effect by the curtain that occurred with the increase/decrease in the accretion rate onto the WD. The asymmetric pulse profile and the shift of the light maximum of the lower-energy pulse to later phases may be caused by the distorted structure of the AC and the curtain due to the inner disk edge rotating faster than the WD during outburst (see sections \ref{sec:res-spin-pulse} and \ref{sec:discuss-spin-pulse}).

	\item To explain almost no energy dependence of high-energy spin pulses at the end of the outburst, we need to introduce the change in the electron scattering effective at $>$10~keV and/or the change in the visible area of the AC to the observer between the on-pulse and off-pulse phases (see sections \ref{sec:res-spin-pulse} and \ref{sec:discuss-spin-pulse}).

	\item QPO signals with a 5699-s period were detected from optical light curves. Some kinds of vertically-thick structures on the disk surface might be associated with this signal. We may observe periodic changes in the irradiated surface area rotating with the Kepler period (see sections \ref{sec:period} and \ref{sec:discuss-qpo}).

\end{itemize}

Our works addressed the mechanism of unique outbursts, the nature of the X-ray emission, and the origin of the WD spin pulse and its time evolution and energy dependence, and the source of optical QPOs in GK Per.
However, there are still unsolved problems.
In particular, the source of the neon, magnesium, iron, and silicon emission lines and the QPO origin are unclear.
To approach their nature, we have to perform some simulations of photoionization and obtain the long-term and high-time-cadence light curves during outbursts in the future.
Also, obtaining high-resolution spectra and detailed spectral analyses are essential.

\begin{acknowledgments}
We are thankful to many amateur observers for providing a lot of data used in this research.
This research was partially supported by the Optical
and Infrared Synergetic Telescopes for Education and Research
(OISTER) program funded by the Ministry of Education, Culture,
Sports, Science and Technology (MEXT) of Japan.
This work was financially supported by Japan Society for the Promotion of Science Grants-in-Aid for Scientific Research (KAKENHI) Grant Numbers JP20K22374 (MK), JP21K13970 (MK), JP21H04491 (SS, MK), and JP24K00673 (WI). 
\end{acknowledgments}

\bibliography{apj-jour,/Users/mariko/cvs}
\bibliographystyle{aasjournal}



\end{document}